%% file: IMBH_paper.tex
\documentclass[aps,prd,twocolumn,groupedaddress,nofootinbib,superscriptaddress]{revtex4-1}
\pdfoutput=1

\usepackage{natbib}
\usepackage{graphicx}
\usepackage{color}
\usepackage{amsmath, amsfonts, amssymb}
\usepackage{hyperref}
\usepackage{subfigure}
\usepackage{ulem}

\begin{document}

 \title{Search for gravitational radiation from intermediate mass black hole binaries in data from the second LIGO-Virgo joint science run}
 
 \input{LSC_Aug2013_Virgo_Nov2013-prd}

 \input{abstract} 

 \maketitle

 \input{introduction}  
 \input{analysis_overview} 
 \input{results}
 \input{discussion}
 \input{acknowledgments}

 \appendix
 \input{appendix_A}

 \bibliography{IMBH_paper}

\end{document}

%% file: LSC_Aug2013_Virgo_Nov2013-prd.tex
\author{%
J.~Aasi$^{1}$,
B.~P.~Abbott$^{1}$,
R.~Abbott$^{1}$,
T.~Abbott$^{2}$,
M.~R.~Abernathy$^{1}$,
T.~Accadia$^{3}$,
F.~Acernese$^{4,5}$,
K.~Ackley$^{6}$,
C.~Adams$^{7}$,
T.~Adams$^{8}$
% ...
}\noaffiliation

\collaboration{LIGO Scientific Collaboration and Virgo Collaboration}

\affiliation {LIGO - California Institute of Technology, Pasadena, CA 91125, USA }% {CIT} {1}
\affiliation {Louisiana State University, Baton Rouge, LA 70803, USA }% {LSU} {2}
\affiliation {Laboratoire d'Annecy-le-Vieux de Physique des Particules (LAPP), Universit\'e de Savoie, CNRS/IN2P3, F-74941 Annecy-le-Vieux, France }% {virgo1} {3}
\affiliation {INFN, Sezione di Napoli, Complesso Universitario di Monte S.Angelo, I-80126 Napoli, Italy }% {virgo2} {4}
\affiliation {Universit\`a di Salerno, Fisciano, I-84084 Salerno, Italy }% {virgo3} {5}
\affiliation {University of Florida, Gainesville, FL 32611, USA }% {UFlorida} {6}
\affiliation {LIGO - Livingston Observatory, Livingston, LA 70754, USA }% {LLO} {7}
\affiliation {Cardiff University, Cardiff, CF24 3AA, United Kingdom }% {Cardiff} {8}
\affiliation {Albert-Einstein-Institut, Max-Planck-Institut f\"ur Gravitationsphysik, D-30167 Hannover, Germany }% {AEIHannover} {9}
\affiliation {Nikhef, Science Park, 1098 XG Amsterdam, The Netherlands }% {virgo4} {10}
\affiliation {LIGO - Massachusetts Institute of Technology, Cambridge, MA 02139, USA }% {MIT} {11}
\affiliation {Instituto Nacional de Pesquisas Espaciais, 12227-010 - S\~{a}o Jos\'{e} dos Campos, SP, Brazil }% {GWINPE} {12}
\affiliation {Inter-University Centre for Astronomy and Astrophysics, Pune - 411007, India }% {IUCAA} {13}
\affiliation {Tata Institute for Fundamental Research, Mumbai 400005, India }% {TIFR} {14}
\affiliation {Syracuse University, Syracuse, NY 13244, USA }% {Syracuse} {15}
\affiliation {University of Wisconsin--Milwaukee, Milwaukee, WI 53201, USA }% {UWM} {16}
\affiliation {Leibniz Universit\"at Hannover, D-30167 Hannover, Germany }% {Leibniz} {17}
\affiliation {INFN, Sezione di Pisa, I-56127 Pisa, Italy }% {virgo5} {18}
\affiliation {Universit\`a di Siena, I-53100 Siena, Italy }% {virgo6} {19}
\affiliation {Stanford University, Stanford, CA 94305, USA }% {Stanford} {20}
\affiliation {The University of Mississippi, University, MS 38677, USA }% {UMiss} {21}
\affiliation {California State University Fullerton, Fullerton, CA 92831, USA }% {Fullerton} {22}
\affiliation {INFN, Sezione di Roma, I-00185 Roma, Italy }% {virgo7} {23}
\affiliation {University of Birmingham, Birmingham, B15 2TT, United Kingdom }% {Birmingham} {24}
\affiliation {Albert-Einstein-Institut, Max-Planck-Institut f\"ur Gravitationsphysik, D-14476 Golm, Germany }% {AEIGolm} {25}
\affiliation {Montana State University, Bozeman, MT 59717, USA }% {MontanaState} {26}
\affiliation {European Gravitational Observatory (EGO), I-56021 Cascina, Pisa, Italy }% {virgo8} {27}
\affiliation {LIGO - Hanford Observatory, Richland, WA 99352, USA }% {LHO} {28}
\affiliation {SUPA, University of Glasgow, Glasgow, G12 8QQ, United Kingdom }% {Glasgow} {29}
\affiliation {APC, AstroParticule et Cosmologie, Universit\'e Paris Diderot, CNRS/IN2P3, CEA/Irfu, Observatoire de Paris, Sorbonne Paris Cit\'e, 10, rue Alice Domon et L\'eonie Duquet, F-75205 Paris Cedex 13, France }% {virgo9} {30}
\affiliation {Columbia University, New York, NY 10027, USA }% {Columbia} {31}
\affiliation {Universit\`a di Pisa, I-56127 Pisa, Italy }% {virgo10} {32}
\affiliation {CAMK-PAN, 00-716 Warsaw, Poland }% {virgo11} {33}
\affiliation {Astronomical Observatory Warsaw University, 00-478 Warsaw, Poland }% {virgo12} {34}
\affiliation {INFN, Sezione di Genova, I-16146 Genova, Italy }% {virgo13} {35}
\affiliation {Universit\`a degli Studi di Genova, I-16146 Genova, Italy }% {virgo14} {36}
\affiliation {San Jose State University, San Jose, CA 95192, USA }% {SanJoseState} {37}
\affiliation {Faculty of Physics, Lomonosov Moscow State University, Moscow 119991, Russia }% {MoscowState} {38}
\affiliation {LAL, Universit\'e Paris-Sud, IN2P3/CNRS, F-91898 Orsay, France }% {virgo15} {39}
\affiliation {NASA/Goddard Space Flight Center, Greenbelt, MD 20771, USA }% {Goddard} {40}
\affiliation {University of Western Australia, Crawley, WA 6009, Australia }% {UWesternAustralia} {41}
\affiliation {Department of Astrophysics/IMAPP, Radboud University Nijmegen, P.O. Box 9010, 6500 GL Nijmegen, The Netherlands }% {virgo16} {42}
\affiliation {Universit\'e Nice-Sophia-Antipolis, CNRS, Observatoire de la C\^ote d'Azur, F-06304 Nice, France }% {virgo17} {43}
\affiliation {Institut de Physique de Rennes, CNRS, Universit\'e de Rennes 1, F-35042 Rennes, France }% {virgo18} {44}
\affiliation {Washington State University, Pullman, WA 99164, USA }% {WashingtonState} {45}
\affiliation {INFN, Sezione di Perugia, I-06123 Perugia, Italy }% {virgo19} {46}
\affiliation {INFN, Sezione di Firenze, I-50019 Sesto Fiorentino, Firenze, Italy }% {virgo20} {47}
\affiliation {Universit\`a degli Studi di Urbino 'Carlo Bo', I-61029 Urbino, Italy }% {virgo21} {48}
\affiliation {University of Oregon, Eugene, OR 97403, USA }% {UOregon} {49}
\affiliation {Laboratoire Kastler Brossel, ENS, CNRS, UPMC, Universit\'e Pierre et Marie Curie, F-75005 Paris, France }% {virgo22} {50}
\affiliation {VU University Amsterdam, 1081 HV Amsterdam, The Netherlands }% {virgo23} {51}
\affiliation {University of Maryland, College Park, MD 20742, USA }% {UMaryland} {52}
\affiliation {University of Massachusetts - Amherst, Amherst, MA 01003, USA }% {UMAmherst} {53}
\affiliation {Laboratoire des Mat\'eriaux Avanc\'es (LMA), IN2P3/CNRS, Universit\'e de Lyon, F-69622 Villeurbanne, Lyon, France }% {virgo24} {54}
\affiliation {Universitat de les Illes Balears, E-07122 Palma de Mallorca, Spain }% {BalearicIslands} {55}
\affiliation {Universit\`a di Napoli 'Federico II', Complesso Universitario di Monte S.Angelo, I-80126 Napoli, Italy }% {virgo25} {56}
\affiliation {Canadian Institute for Theoretical Astrophysics, University of Toronto, Toronto, Ontario, M5S 3H8, Canada }% {CITA-PI} {57}
\affiliation {Tsinghua University, Beijing 100084, China }% {Tsinghua} {58}
\affiliation {University of Michigan, Ann Arbor, MI 48109, USA }% {UMichigan} {59}
\affiliation {Rochester Institute of Technology, Rochester, NY 14623, USA }% {RIT} {60}
\affiliation {INFN, Sezione di Roma Tor Vergata, I-00133 Roma, Italy }% {virgo26} {61}
\affiliation {National Tsing Hua University, Hsinchu Taiwan 300 }% {NTHU} {62}
\affiliation {Charles Sturt University, Wagga Wagga, NSW 2678, Australia }% {CharlesSturt} {63}
\affiliation {Caltech-CaRT, Pasadena, CA 91125, USA }% {CaRT} {64}
\affiliation {Pusan National University, Busan 609-735, Korea }% {PusanNationalU} {65}
\affiliation {Australian National University, Canberra, ACT 0200, Australia }% {ANU} {66}
\affiliation {Carleton College, Northfield, MN 55057, USA }% {Carleton} {67}
\affiliation {INFN, Gran Sasso Science Institute, I-67100 L'Aquila, Italy }% {virgo27} {68}
\affiliation {Universit\`a di Roma Tor Vergata, I-00133 Roma, Italy }% {virgo28} {69}
\affiliation {Universit\`a di Roma 'La Sapienza', I-00185 Roma, Italy }% {virgo29} {70}
\affiliation {University of Brussels, Brussels 1050 Belgium }% {ULB} {71}
\affiliation {Sonoma State University, Rohnert Park, CA 94928, USA }% {SonomaState} {72}
\affiliation {Embry-Riddle Aeronautical University, Prescott, AZ 86301, USA }% {EmbryRiddle} {73}
\affiliation {The George Washington University, Washington, DC 20052, USA }% {GWU} {74}
\affiliation {University of Cambridge, Cambridge, CB2 1TN, United Kingdom }% {Cambridge} {75}
\affiliation {Northwestern University, Evanston, IL 60208, USA }% {Northwestern} {76}
\affiliation {University of Minnesota, Minneapolis, MN 55455, USA }% {UMinnesota} {77}
\affiliation {The University of Texas at Brownsville, Brownsville, TX 78520, USA }% {UTBrownsville} {78}
\affiliation {The University of Sheffield, Sheffield S10 2TN, United Kingdom }% {USheffield} {79}
\affiliation {Wigner RCP, RMKI, H-1121 Budapest, Konkoly Thege Mikl\'os \'ut 29-33, Hungary }% {virgo30} {80}
\affiliation {University of Sannio at Benevento, I-82100 Benevento, Italy }% {USannio} {81}
\affiliation {INFN, Gruppo Collegato di Trento, I-38050 Povo, Trento, Italy }% {virgo31} {82}
\affiliation {Universit\`a di Trento, I-38050 Povo, Trento, Italy }% {virgo32} {83}
\affiliation {Montclair State University, Montclair, NJ 07043, USA }% {MontclairState} {84}
\affiliation {The Pennsylvania State University, University Park, PA 16802, USA }% {PennState} {85}
\affiliation {MTA E\"otv\"os University, `Lendulet' A. R. G., Budapest 1117, Hungary }% {Eotvos} {86}
\affiliation {Universit\`a di Perugia, I-06123 Perugia, Italy }% {virgo33} {87}
\affiliation {Rutherford Appleton Laboratory, HSIC, Chilton, Didcot, Oxon, OX11 0QX, United Kingdom }% {RAL} {88}
\affiliation {Perimeter Institute for Theoretical Physics, Ontario, N2L 2Y5, Canada }% {Perimeter} {89}
\affiliation {American University, Washington, DC 20016, USA }% {American} {90}
\affiliation {University of Adelaide, Adelaide, SA 5005, Australia }% {UAdelaide} {91}
\affiliation {Raman Research Institute, Bangalore, Karnataka 560080, India }% {RRI} {92}
\affiliation {Korea Institute of Science and Technology Information, Daejeon 305-806, Korea }% {KISTI} {93}
\affiliation {Bia{\l }ystok University, 15-424 Bia{\l }ystok, Poland }% {virgo34} {94}
\affiliation {University of Southampton, Southampton, SO17 1BJ, United Kingdom }% {Southampton} {95}
\affiliation {IISER-TVM, CET Campus, Trivandrum Kerala 695016, India }% {IISER-TVM} {96}
\affiliation {Institute of Applied Physics, Nizhny Novgorod, 603950, Russia }% {NizhnyNovgorod} {97}
\affiliation {Seoul National University, Seoul 151-742, Korea }% {SeoulNationalU} {98}
\affiliation {Hanyang University, Seoul 133-791, Korea }% {HanyangU} {99}
\affiliation {IM-PAN, 00-956 Warsaw, Poland }% {virgo35} {100}
\affiliation {NCBJ, 05-400 \'Swierk-Otwock, Poland }% {virgo36} {101}
\affiliation {Institute for Plasma Research, Bhat, Gandhinagar 382428, India }% {IPR-Bhat} {102}
\affiliation {The University of Melbourne, Parkville, VIC 3010, Australia }% {UMelbourne} {103}
\affiliation {INFN, Sezione di Padova, I-35131 Padova, Italy }% {virgo37} {104}
\affiliation {Monash University, Victoria 3800, Australia }% {MonashU} {105}
\affiliation {SUPA, University of Strathclyde, Glasgow, G1 1XQ, United Kingdom }% {Strathclyde} {106}
\affiliation {ESPCI, CNRS, F-75005 Paris, France }% {virgo38} {107}
\affiliation {Argentinian Gravitational Wave Group, Cordoba Cordoba 5000, Argentina }% {AGWG} {108}
\affiliation {Universit\`a di Camerino, Dipartimento di Fisica, I-62032 Camerino, Italy }% {virgo39} {109}
\affiliation {The University of Texas at Austin, Austin, TX 78712, USA }% {UTAustin} {110}
\affiliation {Southern University and A\&M College, Baton Rouge, LA 70813, USA }% {SouthernU} {111}
\affiliation {College of William and Mary, Williamsburg, VA 23187, USA }% {CWM} {112}
\affiliation {IISER-Kolkata, Mohanpur, West Bengal 741252, India }% {IISER-KOL} {113}
\affiliation {National Institute for Mathematical Sciences, Daejeon 305-390, Korea }% {NIMS} {114}
\affiliation {Hobart and William Smith Colleges, Geneva, NY 14456, USA }% {HobartWilliamSmith} {115}
\affiliation {RRCAT, Indore MP 452013, India }% {RRCAT} {116}
\affiliation {SUPA, University of the West of Scotland, Paisley, PA1 2BE, United Kingdom }% {UWS} {117}
\affiliation {Institute of Astronomy, 65-265 Zielona G\'ora, Poland }% {virgo40} {118}
\affiliation {Indian Institute of Technology, Gandhinagar Ahmedabad Gujarat 382424, India }% {IITGN} {119}
\affiliation {Instituto de F\'\i sica Te\'orica, Univ. Estadual Paulista/International Center for Theoretical Physics-South American Institue for Research, S\~ao Paulo SP 01140-070, Brazil }% {ICTP-SAIFR} {120}
\affiliation {Andrews University, Berrien Springs, MI 49104, USA }% {Andrews} {121}
\affiliation {Trinity University, San Antonio, TX 78212, USA }% {Trinity} {122}
\affiliation {University of Washington, Seattle, WA 98195, USA }% {UWashGravity} {123}
\affiliation {Southeastern Louisiana University, Hammond, LA 70402, USA }% {SLU} {124}
\affiliation {Abilene Christian University, Abilene, TX 79699, USA }% {ACU} {125}

%% file: abstract.tex
\begin{abstract}
This paper reports on an unmodeled, all-sky search for gravitational waves from merging intermediate mass black hole binaries (IMBHB). The search was performed on data from the second joint science run of the LIGO and Virgo detectors (July 2009 - October 2010) and was sensitive to IMBHBs with a range up to $\sim 200$ Mpc, averaged over the possible sky positions and inclinations of the binaries with respect to the line of sight. No significant candidate was found. Upper limits on the coalescence-rate density of nonspinning IMBHBs with total masses between 100 and $450 \ \mbox{M}_{\odot}$ and mass ratios between $0.25$ and $1\,$ were placed by combining this analysis with an analogous search performed on data from the first LIGO-Virgo joint science run (November 2005 - October 2007). The most stringent limit was set for systems consisting of two $88 \ \mbox{M}_{\odot}$ black holes and is equal to $0.12 \ \mbox{Mpc}^{-3} \ \mbox{Myr}^{-1}$ at the $90\%$ confidence level. This paper also presents the first estimate, for the case of an unmodeled analysis, of the impact on the search range of IMBHB spin configurations: the visible volume for IMBHBs with nonspinning components is roughly doubled for a population of IMBHBs with spins aligned with the binary's orbital angular momentum and uniformly distributed in the dimensionless spin parameter up to 0.8, whereas an analogous population with antialigned spins decreases the visible volume by $\sim 20\%\,$.
\end{abstract}

%% file: introduction.tex
\section{Introduction} \label{introduction}
Intermediate mass black holes are thought to populate the mass range between few tens of solar masses and $\sim 10^5 \ \mbox{M}_{\odot}$ \cite{Coleman_Miller1}. Although no conclusive detections have been made to date, intermediate mass black holes are very intriguing astrophysical objects, with growing observational and theoretical evidence for their existence  \cite{Coleman_Miller1, van_der_Marel}. Their discovery would be a major breakthrough in our understanding of massive black hole formation \cite{Sesana, Volonteri1}, stellar-cluster evolution \cite{Bash, Ferraro, Gebhardt, Noyola, Trenti, van_den_Bosch} and hyper/ultraluminous x-ray sources \cite{Farrell, Farrell2, Kaaret, Kajava, Strohmayer1, Strohmayer2, Vierdayanti}. Coalescing intermediate mass black hole binaries (IMBHBs) are also the strongest candidate gravitational-wave (GW) sources accessible to ground-based interferometric detectors such as LIGO and Virgo \cite{Abbott1, Accadia}. 

LIGO-Virgo black hole binary searches have focused on the total-mass spectrum below $\sim 450 \ \text{M}_{\odot}$. The observation of more massive systems is penalized by the steep increase of the noise power limiting the detectors' sensitivity at frequencies below $\sim 40$ Hz. The searches have been performed mainly by matched-filtering the data with different families of templates, representing various combinations of the inspiral, merger, and ringdown portions of the waveform \cite{Abbott2, Abbott3, Abadie1, Abadie2, Abadie3, Aasi1, Abbott4}. A further analysis, targeting systems more massive than $50 \ \text{M}_{\odot}$ with ringdown-only templates, has been recently performed in the latest LIGO-Virgo data \cite{Aasi4}.

Black hole binaries have also been searched for with unmodeled methods. In this approach, developed to target GWs shorter than a few seconds, only generic assumptions are made on the signal properties, such as the time duration and frequency range, and events are identified from energy excess with respect to the noise level \cite{Anderson}. Unmodeled and template-based methods share comparable sensitivity when the portion of the signal emitted within the detectors' bandwidth is well localized in the time-frequency domain. At the sensitivity achieved by LIGO and Virgo in the past years, this is the case for black hole binaries more massive than $\sim 100 \ \text{M}_{\odot}$, as shown by a recent LIGO-Virgo study comparing the performances of different black hole binary search methodologies \cite{Mohapatra}. Finally, unmodeled methods have the advantage of not requiring accurate knowledge of the waveform, which helps when reliable models of the targeted signal are not available. Work to develop accurate analytical models of the waveforms including inspiral, merger, and ringdown phases for systems with arbitrarily spinning and precessing companions is still in progress \cite{Hannam2,Hannam:2013waveform,Pan2,Taracchini:2013,Sturani}.

The first unmodeled search for IMBHBs was performed with the coherent WaveBurst algorithm \cite{Klimenko1} on data collected during the fifth LIGO science run (S5, between November 2005 and October 2007) and the first Virgo science run (VSR1, between May and October 2007) \cite{Abadie4}. No significant candidate was found, and upper limits on the coalescence-rate density of nonspinning IMBHBs were calculated for systems with total masses between 100 and 450 $\mbox{M}_{\odot}$ and mass ratios between $0.25$ and $1\,$.

This paper presents the extension of the unmodeled S5-VSR1 IMBHB search to data collected between July 2009 and October 2010 during the sixth LIGO science run (S6) and the second and third Virgo science run (VSR2 and VSR3). The same search algorithm and statistical approach are used as in \cite{Abadie4}, apart from the treatment of uncertainties in upper limits which is discussed in Appendix \ref{appendix_A}$\,$. To estimate the sensitivity of unmodeled searches to light IMBHBs, the astrophysical interpretation of the result was extended to the total-mass range between $50$ and $100 \ \text{M}_{\odot} \,$. The tested IMBHB parameter space was also extended to include companion spins aligned and antialigned with the binary's orbital angular momentum. We expect that the unmodeled nature of the search would allow us to capture precessing systems as well. However, we do not quote upper limits for precessing systems as the sensitivity of the search to GWs from such systems could not be measured because of a lack of accurate waveform models.
 
The paper is organized as follows: Sec. \ref{analysis overview} reports an overview of the analysis, Sec. \ref{results} presents the results of the search, and the results are discussed in Sec. \ref{discussion}$\,$.

%% file: analysis_overview.tex
\section{Analysis overview} \label{analysis overview}

\subsection{Data set} \label{data_set}
The LIGO and Virgo gravitational-wave detectors are kilometer-scale, power-recycled Michelson interferometers with orthogonal Fabry-Perot arms \cite{Abbott1, Accadia}. The LIGO detectors are located at Livingston, Louisiana (L1), and Hanford, Washington (H1). At the Hanford site, a second interferometer (H2) was in operation until 2008 and thus did not contribute to the S6-VSR2/3 science run. The Virgo observatory is in Cascina, Italy (V1). The detectors are currently undergoing upgrades to their advanced configuration, see Sec. \ref{discussion}$ \,$.

The LIGO-Virgo S6-VSR2/3 joint science run is conventionally divided into four epochs: S6a-VSR2 (from July 2009 to September 2009), S6b-VSR2 (from September 2009 to January 2010), S6c (from January 2010 to June 2010) and S6d-VSR3 (from June 2010 to October 2010). The Virgo observatory was not in operation during S6c and only LIGO data is available. Due to maintenance and upgrade work at the detectors, the sensitivities of the instruments varied across the epochs. 

The present analysis was performed with the two networks which preliminary studies showed to have the highest sensitivity: the H1L1V1 and H1L1 configurations. For comparison, the S5-VSR1 search was conducted with the fourfold H1H2L1V1 and the threefold H1H2L1 configurations \cite{Abadie4}. The H1L1 network was analyzed only in times when V1 was not operating; we refer to this as the exclusive H1L1 configuration.

\begin{table}[t!]
\begin{center}
  \begin{tabular}{ccc}
     \hline 
     \hline
     & \multicolumn{2}{c}{Observation time (days)} \ \\
     \cline{2-3} 
     \ Epoch \ \ \ & \ \ \ H1L1V1 \ \ \ & \ \ \ H1L1 \ \\ 
     \hline
     \hline
     \ S6a-VSR2 \ \ \ & \ \ \ 9.0 \ \ \ & \ \ \ 1.6 \ \\
     \ S6b-VSR2 \ \ \ & \ \ \ 15.1 \ \ \ & \ \ \ 7.3 \ \\ 
     \ S6c \ \ \ & \ \ \ - \ \ \ & \ \ \ 48.2 \ \\ 
     \ S6d-VSR3 \ \ \ & \ \ \ 18.0 \ \ \ & \ \ \ 22.1 \ \\ 
     \hline
     \hline
     \ Total \ \ \ & \ \ \ 42.1 \ \ \ & \ \ \ 79.2 \ \\ 
     \hline
     \hline
  \end{tabular}
\end{center}
\caption{The H1L1V1 and exclusive H1L1 observation time analyzed for each of the S6-VSR2/3 epochs. The values denote the observation times collected after the application of all vetoes used in this search.}
\label{observation_time}
\end{table}

To remove from the analysis data segments likely to be significantly affected by nonstationary noise sources, the data was selected based on data-quality vetoes \cite{Slutsky, Aasi3, McIver}. A combination of site activity and the absence of the fourth detector available during S5-VSR1, H2, meant that a higher rate of nonstationary noise events was observed during the S6-VSR2/3 run\footnote{Due to the lack of signal constraints, unmodeled algorithms are more efficient at rejecting noise events when the search is conducted on large networks.}$\,$. Therefore, a broader class of vetoes was applied in this search than in the S5-VSR1 analysis. This reduced the total H1L1V1 and exclusive H1L1 observation time available after the application of the class of vetoes used during the S5-VSR1 search by a further $\sim 20 \%$ and $\sim 8 \%$, respectively. The total observation time analyzed for this search is reported in Table \ref{observation_time}$\,$.

We also applied event-by-event vetoes based on instrumental and environmental measurements \cite{Smith}, with typical duration $< 1 \ \text{s}$.  This did not significantly reduce the available observation time.

\subsection{Data-analysis algorithm}
The coherent WaveBurst algorithm was developed within the LIGO and Virgo Collaborations for coherent, unmodeled GW-burst searches in data from networks of arbitrarily aligned detectors \cite{Klimenko1}. The search is conducted on a time-frequency representation of the data \cite{Klimenko2}. A constrained maximum-likelihood approach is used to identify coherent network events from the time-frequency regions with energy excess relative to the noise level and assign them a number of coherent statistics \cite{Klimenko1, Klimenko3, Klimenko4}. The constraints are introduced to suppress unphysical solutions. For compact-binary searches, the reconstruction of elliptically polarized events is enforced to improve the rejection of noise events \cite{Pankow1}.

The reconstructed events are selected by applying cuts on three major coherent statistics: the network correlation coefficient ($cc$) and the network energy disbalance ($\lambda$), which estimate the overall consistency of the events, and the coherent network amplitude ($\eta$), which estimates the event strength \cite{Klimenko1, Pankow1, Abadie4}. In this search, only events reconstructed with $cc > 0.7$ and $\lambda < 0.4$ were considered for further follow-up.

\subsection{Background estimation} \label{background_estimation} 
The background for this search was empirically estimated by analyzing a few hundred independent time-shifted data sets. Since noise is assumed to be uncorrelated between sites, introducing relative time delays larger than the GW travel time  ($\lesssim 30$ ms between the LIGO and Virgo facilities) is an effective way to generate an instance of the accidental background. The effective H1L1V1 and H1L1 background livetime accumulated for each S6-VSR2/3 epoch is reported in Table \ref{background_time}$\,$.

Considering a few hundreds time lags enabled an estimate of the tails of the background distribution with the precision of a few percent. This level of accuracy was considered sufficient for an initial estimate of the false alarm probability of the loudest observed events. Additional time lags would have been analyzed had loud GW candidates been identified and a more precise estimate of the background tails had been required.

\begin{table}[t!]
\begin{center}
  \begin{tabular}{ccc}
     \hline 
     \hline
     & \multicolumn{2}{c}{Background livetime (years)} \ \\
     \cline{2-3}
     \ Epoch \ \ \ & \ \ \ H1L1V1 \ \ \ & \ \ \ H1L1 \ \\ 
     \hline
     \hline
     \ S6a-VSR2 \ \ \ & \ \ \ 9.1 \ \ \ & \ \ \ 8.4 \ \\
     \ S6b-VSR2 \ \ \ & \ \ \ 14.7 \ \ \ & \ \ \ 18.4 \ \\ 
     \ S6c \ \ \ & \ \ \ - \ \ \ & \ \ \ 39.4 \ \\ 
     \ S6d-VSR3 \ \ \ & \ \ \ 18.8 \ \ \ & \ \ \ 32.9 \ \\ 
     \hline
     \hline
     \ Total \ \ \ & \ \ \ 42.6 \ \ \ & \ \ \ 99.1 \ \\ 
     \hline
     \hline
  \end{tabular}
\end{center}
\caption{The H1L1V1 and H1L1 background livetime accumulated for each of the S6-VSR2/3 epochs. The search background was estimated on the data segments passing all vetoes used in this search.}
\label{background_time}
\end{table}

\subsection{Search sensitivity} \label{search_sensitivity}
The search sensitivity is quoted as the visible volume for IMBHB mergers. Its calculation relied on Monte Carlo detection-efficiency studies. Simulated signals modeling the gravitational radiation emitted by coalescing IMBHBs were added via software to LIGO-Virgo data and searched for with coherent WaveBurst.

Two waveform families were used: EOBNRv2 \cite{Pan} and IMRPhenomB~\cite{Ajith2}. The EOBNRv2 family models GWs from binaries with nonspinning companions and was used to combine the present analysis with the S5-VSR1 search, whose sensitivity was assessed using this family~\cite{Abadie4}. Only the dominant $(l, m) = (2,2)$ mode was used for both searches. The IMRPhenomB family models companions with spins aligned or antialigned with the binary's orbital angular momentum and was used to estimate the impact of these spin configurations on search sensitivity.

The simulated signals were uniformly distributed over the total-mass spectrum between 50 and 450 $\text{M}_{\odot}$ and mass-ratio range between $0.25$ and  $1 \,$. The IMRPhenomB waveforms were also uniformly distributed over the spin interval $[-0.8, \, 0.8] \,$, the recommended range of validity of IMRPhenomB waveforms \cite{Ajith2}. Here the spin interval is expressed in terms of the dimensionless spin parameter $\chi_{1, \, 2} = S_{1, \, 2} / m_{1, \, 2}^2$, where $S$ and $m$ are the spin angular momentum and the mass of the two binary components. The uniform distributions in total mass, mass ratio and spin were motivated by the lack of astrophysical constraints on IMBHB parameters. The simulated signals were uniformly distributed in volume, polarization angle and binary inclination with respect to the line of sight. The results in the following sections are averaged over the binary's sky position and orientation.

Following the approach used in the S5-VSR1 search \cite{Abadie4}, for a given source population and threshold on $\eta$, the visible volume $V_{\text{vis}}$ was computed as a function of the binary parameters as
\begin{equation} \label{visible_volume}
 V_{\text{vis}} \left(m_1, m_2, \chi_1, \chi_2, \eta \right) = \sum_{\eta_i > \eta } \frac{1}{\rho_i} = \sum_{\eta_i>\eta} \frac {4 \pi r_i^2 } {\frac{dN_{\text{inj}}}{dr} \left(r_i \right) } \ .
\end{equation}
In the above formula, $\rho$ denotes the number density of the simulated signals, injected at distance $r$ with radial density $dN_{\text{inj}}/dr \,$, and the sum runs over the set of injections recovered  with coherent network amplitude $\eta_i$ above $\eta$. The search sensitivity can be equivalently quoted in terms of the search range, calculated as the radius of the sphere with volume $V_{\text{vis}} \,$.

\subsection{The false alarm rate density statistic} \label{False_alarm_density}
The S5-VSR1 and S6-VSR2/3 IMBHB searches were combined using the false alarm rate density (FAD) statistic \cite{Abadie4, Pankow2}, which is defined as
\begin{equation}\label{Eq:FAD}
 \mbox{FAD} \left( \eta \right) = \frac{1}{T_{\text{bkg}}} \sum_{\eta_i > \eta} \frac{1}{ \bar{V}_{\text{vis}} \left( \eta_i \right) } \ .
\end{equation}
Here $T_{\text{bkg}}$ is the effective background livetime, $\bar{V}_{\text{vis}}$ is $V_{\text{vis}}$ averaged over the investigated parameter space and the sum runs over the background events with $\eta_i > \eta \,$. The FAD statistic estimates the rate density of background events above a given threshold.  We rank events by FAD across searches, referring to events with lower FAD values as louder events.

When inverting Eq. (\ref{Eq:FAD}) to obtain the $\eta$ threshold corresponding to a certain FAD, we chose the upper bound on the range of $\eta$ corresponding to this FAD.  This procedure allowed us to obtain a conservative lower estimate of $V_{\text{vis}} \left(\text{FAD} \right)$.

The total time-volume product surveyed by the combined searches was calculated as
\begin{equation} \label{Eq:nu}
 \nu \left(\text{FAD} \right) = \sum_{k} T_{\text{obs}, \, k} \, V_{\text{vis}, \, k} \left(\text{FAD} \right) \ .
\end{equation}
In the above equation, the sum runs over the searches and $T_{\text{obs}, \, k}$ and $V_{\text{vis}, \, k}$ are the observation time and the visible volume of the $k$th search, respectively. The mean number of noise events expected within $\nu \left(\text{FAD} \right)$ is conservatively overestimated as 
\begin{equation} \label{mu_formula}
 \mu (\mbox{FAD}) = \mbox{FAD} \, \times \, \nu \left(\text{FAD} \right) \ .
\end{equation}

The significance of GW candidates is set by the false alarm probability (FAP), which is the probability of observing $N$ or more noise events with a FAD statistic below threshold. For a Poisson distribution of background events with mean $\mu (\mbox{FAD})$, the false alarm probability is
\begin{equation} \label{FAP_formula}
 \mbox{FAP}(N) = 1 - \sum_{n = 0}^{N - 1} \frac{\left[\mu (\text{FAD})\right]^{n}}{n!} \, e^{-\mu \left(\text{FAD} \right)} \ .
\end{equation}
Following the definition of $\mu$(FAD) in Eq. (\ref{mu_formula}), the estimated false alarm probability is designed to be a conservative overestimate. Hereafter, we will refer to the false alarm probability calculated setting $N = 1$ in Eq. (\ref{FAP_formula}) as single-event false alarm probability.

\subsection{Uncertainties on the search range} \label{impact_of_systematics}
Three sources of uncertainty on the search range were considered. In order of relevance, these were: calibration uncertainties of the LIGO and Virgo detectors, waveform systematics and statistical errors.

Calibration uncertainties affect the GW strain reconstructed at the detectors. During the S6-VSR2/3 science run, the largest amplitude uncertainty was $19\%$ at the L1 detector \cite{Bartos, Accadia4, Accadia2}. We conservatively assumed a calibration induced uncertainty on the search range of $19 \%$ for each detector and over the whole S6-VSR2/3 run. Any additional constant calibration uncertainty would potentially affect the upper limits presented in this paper, though not our statements of (non)detection.

Waveform systematics arise from the discrepancy between the considered approximate waveform families and the actual GW signature. The impact on this search was estimated by comparing the optimal matched filter signal-to-noise ratios obtained with EOBNRv2 waveforms ($\text{SNR}_{\text{E}}$) and with numerical models of the same sources ($\text{SNR}_{\text{N}}$). The comparison was based on the quantity 
\begin{equation}
 \Delta = \frac{\text{SNR}_{\text{E}} - \text{SNR}_{\text{N}}}{\text{SNR}_{\text{N}}} \ .
\end{equation}
The $\Delta$ were found to vary within $\left[-8\%, \, 14 \% \right]\,$ over most of the tested parameter space. To account for the waveform systematics, the search ranges calculated with EOBNRv2 waveforms were rescaled upwards (downwards) by a factor $\Delta$ in the regions of the parameter space where $\Delta$ was negative (positive).

The statistical error originates from the finite number of injections performed. This uncertainty on the search range was calculated as in \cite{Abadie4} and was $\lesssim 2 \%$ over most of the investigated parameter space.

%% file: results.tex
\section{Results} \label{results}

\subsection{Loudest events} \label{loudest_event}
\begin{figure*}[htp]
 \centering
 \subfigure[]{\includegraphics{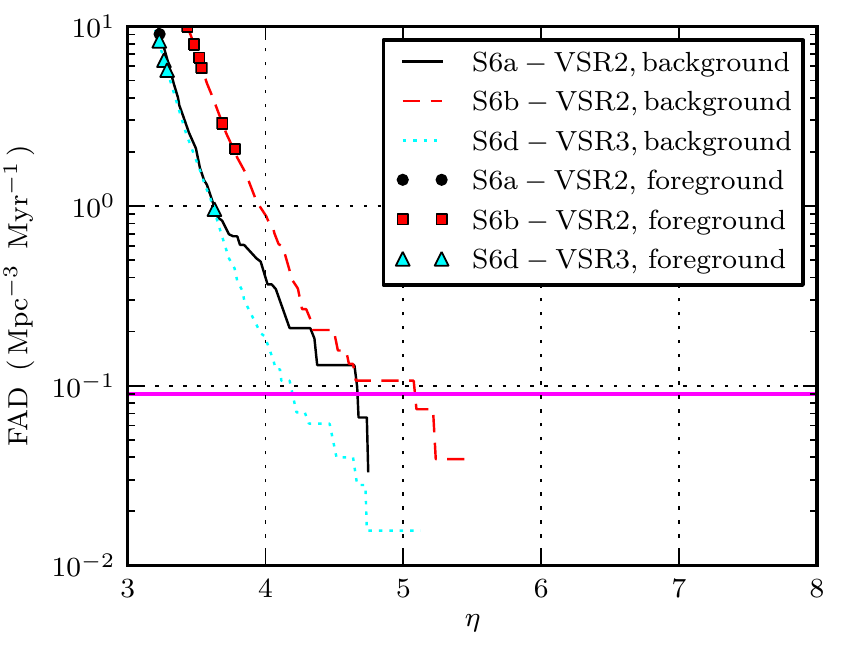}}
 \hspace{3mm}
 \subfigure[]{\includegraphics{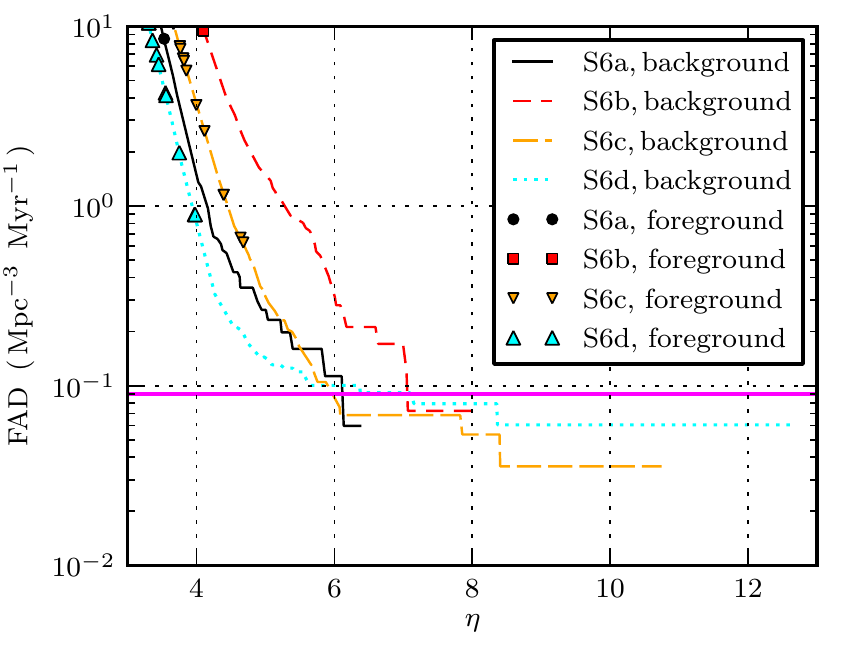}}
  \caption{(color online). (a) H1L1V1 FAD background distributions as a function of $\eta$ and reconstructed events (foreground). (b) H1L1 FAD background distributions as a function of $\eta$ and reconstructed events (foreground). The horizontal line denotes the FAD threshold at which the S5-VSR1 and S6-VSR2/3 searches were combined. The threshold was determined by the loudest event identified by the combined S5-VSR1 and S6-VSR2/3 analyses, which had been reconstructed in H1H2L1V1 S5-VSR1 data.}
 \label{FAD_distribution}
\end{figure*}

None of the events identified by the search (foreground events) or groups of loudest $N$ foreground events were significant enough to claim a GW detection. The foreground events are shown in Fig. \ref{FAD_distribution}$\,$, together with the H1L1V1 and H1L1 FAD background distributions. The different $\bar{V}_{\text{vis}}$ used to construct the FAD distributions were calculated with EOBNRv2 waveforms.

\begin{table*}[htp]
 \begin{center}
   \begin{tabular}{ccccccc}
      \hline 
      \hline
      \ Rank \ & \ FAD $(\text{Mpc}^{-3} \ \text{Myr}^{-1})$ \ & \ FAP \ & \ Global positioning system time \ & \ Network \ & \ Epoch \ & \ $\eta$ \ \\ 
      \hline
      \hline
       \ 1 \ & \ 0.63 \ & \ $44 \%$ \ & \ 951496848 \ & \ H1L1 \ & \ S6c \ & \ 4.7 \ \\
       \ 2 \ & \ 0.67 \ & \ $46 \%$ \ & \ 947225014 \ & \ H1L1 \ & \ S6c \ & \ 4.6 \ \\
       \ 3 \ & \ 0.90 \ & \ $49 \%$ \ & \ 966874796 \ & \ H1L1 \ & \ S6d-VSR3 \ & \ 4.0 \ \\ 
       \ 4 \ & \ 0.90 \ & \ $49 \%$ \ & \ 962561544 \ & \ H1L1 \ & \ S6d-VSR3 \ & \ 4.0 \ \\
       \ 5 \ & \ 0.96 \ & \ $35 \%$ \ & \ 971422542 \ & \ H1L1V1 \ & \ S6d-VSR3 \ & \ 3.6 \ \\
       \hline
       \hline
     \end{tabular}
   \end{center}
  \caption{Loudest events reconstructed by the S6-VSR2/3 search. The events are ranked by the false alarm rate density (FAD) at which they were identified. The false alarm probability (FAP) is calculated as single-event false alarm probability.}
 \label{table_loudest_events}
 \end{table*}

The loudest foreground events, ranked by FAD, are summarized in Table \ref{table_loudest_events}$\,$. The first ranked event was identified in H1L1 S6c data on March 1, 2010, at 16:40:33 UTC time and has a single-event false alarm probability equal to $44 \%$. The event in Table \ref{table_loudest_events} with the lowest single-event false alarm probability, equal to $35 \%$, was reconstructed in H1L1V1 S6d-VSR3 data and is the fifth ranked event. It is not contradictory that louder events could have higher single-event false alarm probabilities: compared to the other events in Table \ref{table_loudest_events}, the considered H1L1V1 event was identified by a search conducted with a different network and with a lower collected observation time.

As a sanity check, follow-up analyses were performed with coherent WaveBurst on the events in Table \ref{table_loudest_events}$\,$. The loudest event was further followed up with a Bayesian parameter estimation algorithm, specifically developed for compact binary systems \cite{S6PECBC}. The tests showed no evidence for an event that stands out above the background. 

As no GWs were detected, the main astrophysical result was the calculation of combined S5-VSR1 and S6-VSR2/3 upper limits on the coalescence-rate density of nonspinning IMBHBs. These were calculated with the loudest event statistic (see Sec. \ref{event_rate_upper_limit}$\,$). The S5-VSR1 and S6-VSR2/3 searches were therefore combined at the FAD threshold set by the loudest event identified by the combined analyses.  This event, reconstructed by the S5-VSR1 search in H1H2L1V1 data, had a  FAD of $0.09 \ \text{Mpc}^{-3} \ \text{Myr}^{-1}$ \cite{Abadie4}. Hereafter, this FAD value will be denoted as $\text{FAD}^*$.

\subsection{Search ranges} \label{search_ranges}
We calculated the search ranges corresponding to the surveyed volumes as a function of the companion masses. The calculation was based on simulation studies conducted with EOBNRv2 waveforms at the $\eta$ thresholds determined by the $\text{FAD}^*$ value. 

\begin{figure*}[htp]
 \centering
 \subfigure[]{\includegraphics{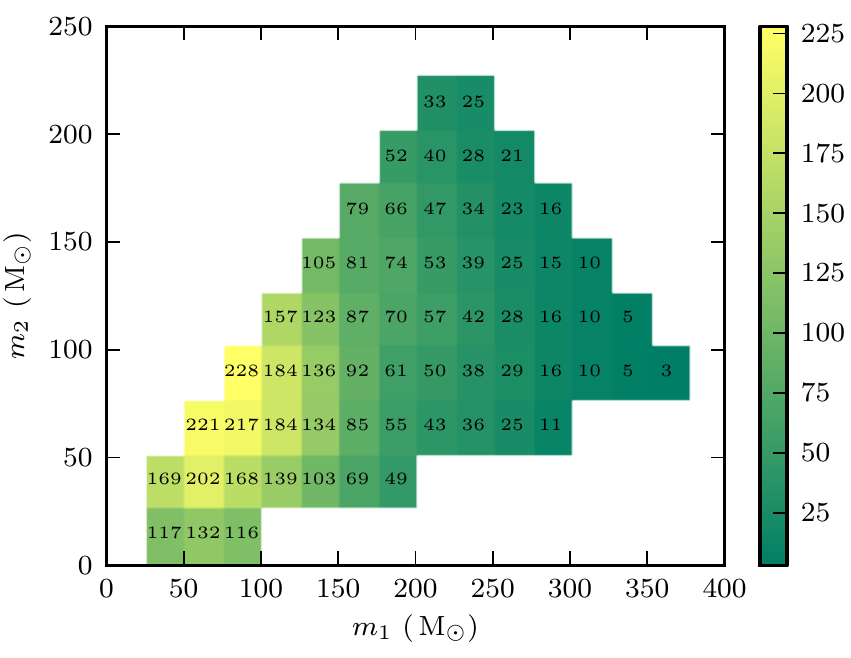}}
 \hspace{3mm} 
 \subfigure[]{\includegraphics{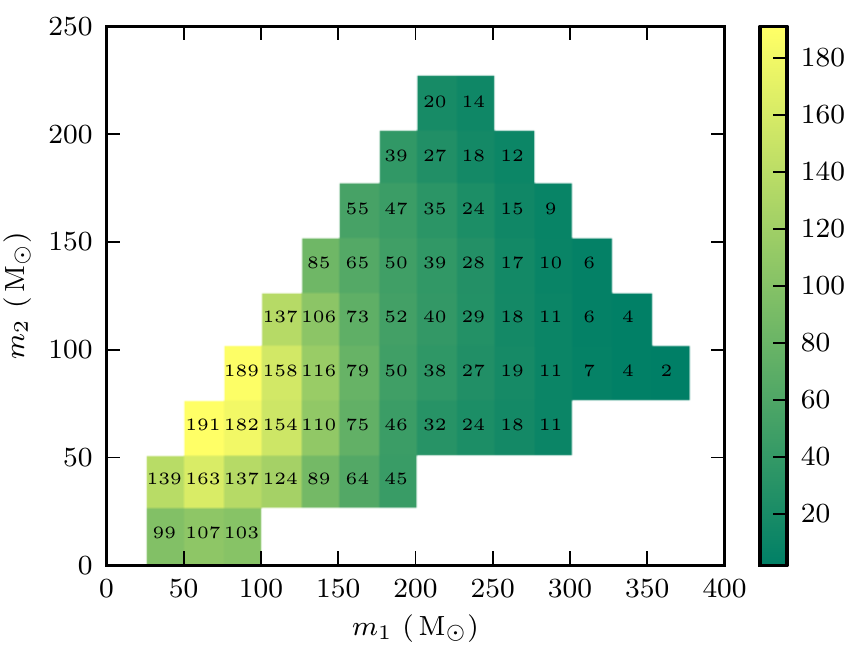}}
   \caption{(color online). (a) S6d-VSR3 H1L1V1 search ranges in Mpc as a function of the companion masses. (b) S6c H1L1 search ranges in Mpc as a function of the companion masses. The ranges were calculated via simulation studies conducted with EOBNRv2 waveforms and are reported as color scales.} 
 \label{search_range}
\end{figure*}

The H1L1V1 and H1L1 largest search ranges were achieved during the S6d-VSR3 and S6c epochs, respectively. The results are reported in Fig. \ref{search_range}$\,$. The H1L1V1 (H1L1) best search range was equal to $\sim 230$ Mpc ($\sim 190$ Mpc) and was calculated in the mass bin centered at $88 + 88 \ \text{M}_{\odot}$ ($63 + 63 \ \text{M}_{\odot}$). Over the other S6-VSR2/3 epochs, the search ranges calculated in the most sensitive mass bin were found to decrease by at most $\sim 30 \%$ for both the H1L1V1 and H1L1 networks.

At the considered FAD threshold, the H1L1V1 and H1L1 search ranges achieved during the S6c and S6d-VSR3 epochs, i.e., over most of the accumulated observation time, were comparable to those of the S5-VSR1 search. The S5-VSR1 H1H2L1V1 and H1H2L1 analyses were sensitive to merging IMBHBs up to $\sim 240$ and $\sim 190$ Mpc, respectively \cite{Abadie4}. 

The S6-VSR2/3 search range was also estimated over the total-mass spectrum between 50 and 100 $\text{M}_{\odot} \,$. This region of the parameter space had not been considered for the S5-VSR1 search. Over the whole S6-VSR2/3 run, the H1L1V1 (H1L1) search range was found to vary between $\sim 75$ and $\sim 170$ Mpc ($\sim 70$ and $\sim 140$ Mpc) in this mass range. 

\subsection{Impact of spins on the search range}
The results in Sec. \ref{search_ranges} were calculated for nonspinning black holes. However, observations suggest that black holes could have significant spin \cite{McClintock, Reynolds}. 
The amount of energy lost to GWs by coalescing binaries depends crucially on the spins of the companions. Compared to the case of nonspinning components, aligned (antialigned) spin configurations increase (decrease) the energy released by the system \cite{Campanelli}.  Monte Carlo simulation studies were conducted with IMRPhenomB waveforms to estimate the impact of aligned and anti-aligned companion spins on the visible volume surveyed by the search. 

\begin{figure}[t!]
 \begin{center} 
 \includegraphics{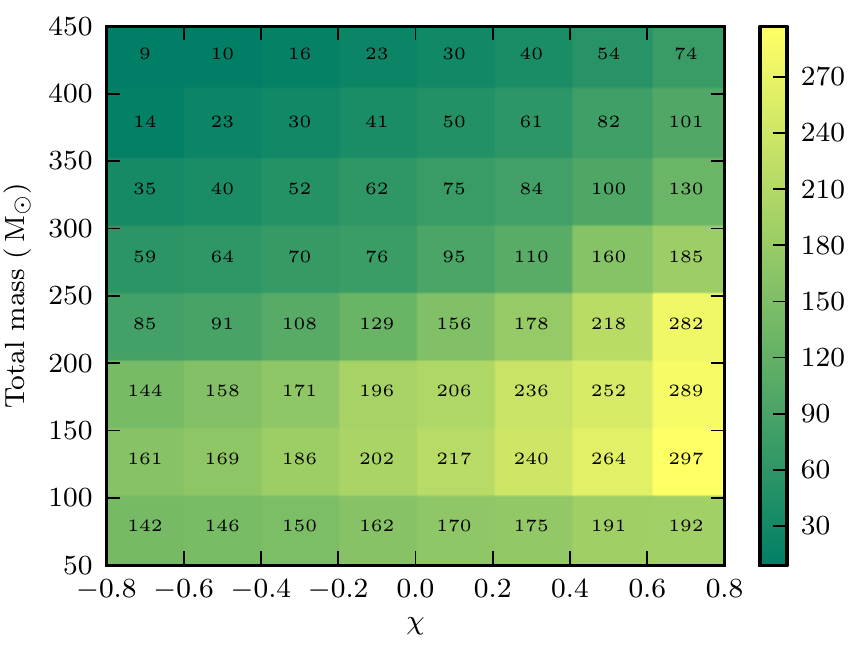}
 \caption{(color online). Search ranges in Mpc as a function of the binary total mass and of the effective spin parameter $\chi$, calculated with IMRPhenomB waveforms on S6d-VSR3 H1L1V1 data. The ranges were averaged over the mass-ratio interval between $0.25$ and $1\,$ and are reported as color scale.}
 \label{search_range_spin}
 \end{center}  
\end{figure}

An example of the impact of spins on the coherent WaveBurst search ranges is shown in Fig. \ref{search_range_spin}$\,$. The ranges are expressed as a function of the binary total mass and of the effective spin parameter $\chi\,$, defined as
\begin{equation}
 \chi = \frac{m_1 \chi_1 + m_2 \chi_2}{m_1 + m_2} \ .
\end{equation}
Note the increase of the search range for progressively larger $\chi$ values at a given total mass. The results in Fig. \ref{search_range_spin} were calculated by averaging over the mass-ratio range between $0.25$ and $1\,$.  A different distribution of mass ratios would not modify the general trend shown in the plot.

For a quantitative estimate of the impact of aligned and antialigned spin configurations on the analysis, we compared the $\bar{V}_{\text{vis}}$ calculated for IMRPhenomB waveforms with and without spins. The cases of aligned and antialigned spins were tested separately as recent studies suggest that aligned configurations could be more likely \cite{Gerosa}. Averaging over the aligned-spin range $0 < \chi_{1, \, 2} < 0.8 \,$, the visible volume $\bar{V}_{\text{vis}}$ was found to be roughly doubled relative to $\bar{V}_{\text{vis}} (\chi_{1, \, 2} = 0)\,$. The visible volume $\bar{V}_{\text{vis}}$ was found to decrease by roughly $-20\%$ when averaging over the anti-aligned spin range $- 0.8 < \chi_{1, \, 2} < 0 \,$. Finally, averaging over the $\chi_{1, \, 2}$ range from $- 0.8$ to $0.8 \,$, the accessible $\bar{V}_{\text{vis}}$ increases by roughly $40\%\,$.

The values above are averaged over assumed uniform spin distributions. However, the spins of intermediate mass black holes may not be distributed uniformly in nature. Relative to nonspinning binaries, sensitive volumes could be more than tripled or less than halved by more extreme aligned or antialigned distributions, respectively.

Aligned and antialigned spin configurations are only a limited class of realistic scenarios. In general, misaligned spin configurations are likely, and these will induce precession. The physics of two precessing black holes orbiting each other in a strongly relativistic regime is challenging \cite{Apostolatos} and dedicated waveforms are currently under development \cite{Hannam2,Hannam:2013waveform,Pan2,Taracchini:2013,Sturani}. The lack of reliable waveforms at the time of the search made it impossible to estimate the search sensitivity to precessing IMBHBs. Nevertheless, precession is not expected to strongly affect the detection efficiency of this search. This is a major advantage shared by unmodeled strategies compared to matched filtering, which could be significantly affected by differences between the targeted GW signal and the considered template family.

\subsection{Rate density upper limits} \label{event_rate_upper_limit}
We placed frequentist upper limits on the coalescence-rate density of nonspinning IMBHBs at the $90 \%$ confidence level. The upper limits were calculated by combining the S5-VSR1 H1H2L1V1 and H1H2L1 searches with the S6-VSR2/3 H1L1V1 and H1L1 analyses. The calculation was performed on the IMBHB parameter space common to the S5-VSR1 and S6-VSR2/3 searches. The upper limits were set for each tested mass bin with the loudest event statistic \cite{Biswas, Sutton}:
\begin{equation} \label{UL_formula}
 \mathcal{R}_{90 \%} = \frac{2.3}{\nu(\text{FAD}^*)} \ .
\end{equation}
In the above equation, $\nu(\text{FAD}^*)$ is the total time-volume product surveyed by the combined searches at the FAD threshold of $0.09 \ \text{Mpc}^{-3} \, \text{Myr}^{-1}$.

\begin{figure}[t]
 \begin{center}  
  \includegraphics{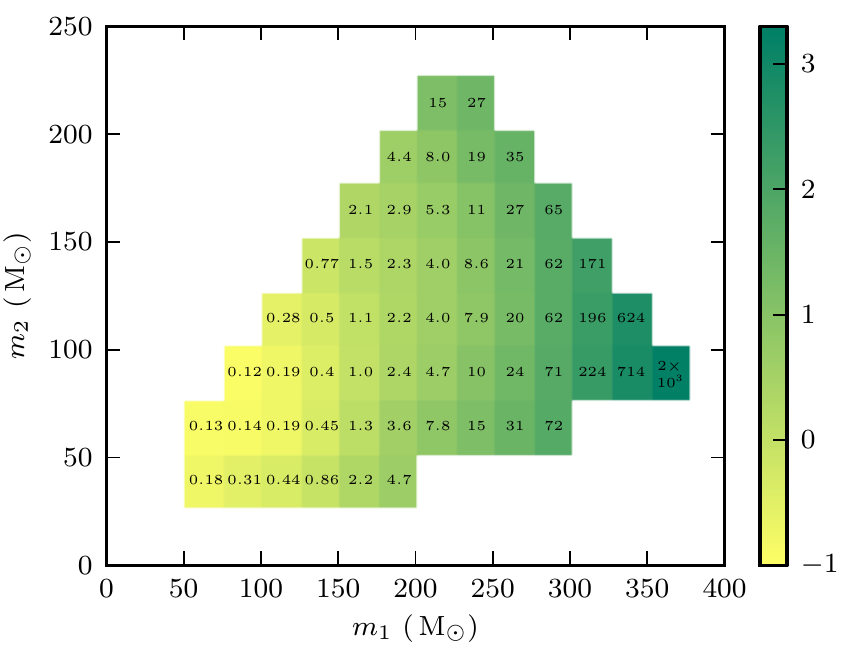}
  \caption{(color online). $90\%$-confidence upper limits in $\text{Mpc}^{-3} \, \text{Myr}^{-1}$ on the merger-rate density of nonspinning IMBHBs as a function of the companion masses. The values were computed with EOBNRv2 waveforms and by combining the S5-VSR1 and S6-VSR2/3 searches. The result includes the uncertainties on the search range discussed in Sec. \ref{impact_of_systematics}$ \,$. The color scale expresses the upper limits as powers of 10$\,$.}
  \label{S5_S6_UL}
 \end{center}  
\end{figure}

The upper limits were conservatively corrected to account for the uncertainties on the search range described in Section \ref{impact_of_systematics}$ \,$. The approach we followed to include the uncertainties differs from the procedure considered for the S5-VSR1 search and is discussed in Appendix \ref{appendix_A}$\,$. The combined upper limits are reported in Fig. \ref{S5_S6_UL}$ \,$. The tightest constraint was placed on the mass bin centered at $88 + 88 \ \mbox{M}_{\odot}$ and is equal to $0.12 \ \text{Mpc}^{-3} \, \text{Myr}^{-1}$.

Astrophysical models suggest globular clusters (GC) as suitable environments for hosting IMBHBs \cite{Coleman_Miller3}. Assuming a GC density of 3 GC $\text{Mpc}^{-3}$ \cite{Portegies2}, we converted the best upper limit to an astrophysical density of $40 \ \text{GC}^{-1} \ \text{Gyr}^{-1} \,$. The result is more than two orders of magnitude away from $0.1 \ \text{GC}^{-1} \, \text{Gyr}^{-1}$, the coalescence-rate density estimated by assuming one IMBHB merger in each GC over the cluster lifetime $(\sim 10 \, \, \text{Gyr})$.

%% file: discussion.tex
\section{Discussion} \label{discussion}
This paper reported on the search for IMBHBs conducted with the coherent WaveBurst algorithm over data collected by the LIGO and Virgo gravitational-wave detectors between July 2009 and October 2010 (S6-VSR2/3 science run). No candidate was identified. Upper limits on the merger-rate density of nonspinning IMBHBs were placed by combining this search with an analogous analysis performed on LIGO-Virgo data collected between November 2005 and October 2007 (S5-VSR1 science run). The most stringent upper limit was set for systems consisting of two $88 \ \text{M}_{\odot}$ companions and is equal to $0.12 \ \text{Mpc}^{-3} \, \text{Myr}^{-1}$ at the $90 \%$ confidence level. 

Although the S5-VSR1 and S6-VSR2/3 searches shared comparable sensitivities, the S5-VSR1 analysis provided the main contribution to the combined upper limits, mostly due to the longer analyzed observation time.  The combined upper limits in Fig. \ref{S5_S6_UL} are comparable to the S5-VSR1 upper limits in \cite{Abadie4} in the most sensitive region of the parameter space and less stringent at high total masses. This is primarily due to the different procedure adopted to conservatively correct the upper limits for the uncertainties on search range.  Furthermore, the decrease in the sensitivity of the LIGO instruments below $\sim 60$ Hz between S5 and S6 reduced the range of the S6-VSR2/3 search for IMBHBs with total mass $\gtrsim 200 \ \mbox{M}_{\odot}$. These issues are discussed further in Appendix \ref{appendix_A}$\,$.

It is worth comparing the upper limits placed by this analysis and by matched-filtering binary black hole  searches conducted on the same data. Due to the complementary total-mass ranges investigated with template-based methods using full inspiral-merger-ringdown waveforms and the unmodeled search described here (below and above $100 \ \text{M}_{\odot}$, respectively), we compare the results set for systems consisting of two $50 \ \text{M}_{\odot}$ nonspinning companions. The upper limit reported in this paper, $0.13 \ \text{Mpc}^{-3} \, \text{Myr}^{-1}\,$, is less stringent than the one offered by the template-based analysis, $0.07 \ \text{Mpc}^{-3} \, \text{Myr}^{-1}$ \cite{Aasi1}. This reflects the increasing power of matched-filtering approaches to distinguish genuine GWs from noise when multiple cycles of the waveform are present over a broad frequency band. However, the comparison must be done with caution due to a number of differences between the two analyses, primarily the statistical approach to computing upper limits and the handling of uncertainties. Finally, the analysis presented here searches over a wider parameter space and is more robust against unmodeled features, such as those arising from strongly precessing signals, which are more likely to be rejected by the matched-filtering search.

Although the combined upper limits presented in this paper do not challenge astrophysical models, the results we report are currently the best constraint on the IMBHB merger-rate density based on direct measurements. Furthermore, the S5-VSR1 and S6-VSR2/3 analyses provide a major benchmark for the IMBHB searches which will be conducted with the second-generation ground-based interferometric detectors. 

The second-generation detectors are the upgraded LIGO and Virgo observatories and the comparably sensitive KAGRA interferometer \cite{Harry, Accadia3, Somiya}. This advanced class of detectors, which will come online in a few years, is expected to significantly increase the sensitivity achieved during the past science runs and to extend the lower-frequency end of the detector bandwidth from $\sim 40$ Hz down to $\sim 10$ Hz \cite{Aasi2}. Simulation studies suggest that coherent WaveBurst analyses conducted with networks of second-generation detectors could be sensitive to IMBHB mergers up to the Gpc scale within the total-mass spectrum below $\sim 1000 \ \mbox{M}_{\odot}$ \cite{Mazzolo2}. Thus, second-generation detectors may commence the era of IMBHB astronomy.

%% file: acknowledgments.tex
\section{Acknowledgments} \label{acknowledgments}
The authors gratefully acknowledge the support of the United States National Science Foundation for the construction and operation of the LIGO Laboratory, the Science and Technology Facilities Council of the United Kingdom, the Max-Planck-Society, and the State of Niedersachsen/Germany for support of the construction and operation of the GEO600 detector, and the Italian Istituto Nazionale di Fisica Nucleare and the French Centre National de la Recherche Scientifique for the construction and operation of the Virgo detector. The authors also gratefully acknowledge the support of the research by these agencies and by the Australian Research Council, the International Science Linkages program of the Commonwealth of Australia, the Council of Scientific and Industrial Research of India, the Istituto Nazionale di Fisica Nucleare of Italy, the Spanish Ministerio de Econom\'ia y Competitividad, the Conselleria d'Economia Hisenda i Innovaci\'o of the Govern de les Illes Balears, the Foundation for Fundamental Research on Matter supported by the Netherlands Organisation for Scientific Research, the Polish Ministry of Science and Higher Education, the FOCUS Programme of Foundation for Polish Science, the Royal Society, the Scottish Funding Council, the Scottish Universities Physics Alliance, the National Aeronautics and Space Administration, OTKA of Hungary,
the Lyon Institute of Origins (LIO), the National Research Foundation of Korea, Industry Canada and the Province of Ontario through the Ministry of Economic Development and Innovation, the National Science and Engineering Research Council Canada, the Carnegie Trust, the Leverhulme Trust, the David and Lucile Packard Foundation, the Research Corporation, and the Alfred P. Sloan Foundation. This document was assigned the LIGO number LIGO-P1300158.

%% file: appendix_A.tex
\section{THE S5-VSR1 AND S6-VSR2/3 UPPER LIMITS} \label{appendix_A}
The upper limits in Figure \ref{S5_S6_UL} were calculated by combining the constraints set on S5-VSR1 and S6-VSR2/3 data. The S5-VSR1 and S6-VSR2/3 upper limits were computed by combining the H1H2L1V1 and H1H2L1 analyses and the H1L1V1 and H1L1 searches, respectively. The S5-VSR1 and S6-VSR2/3 upper limits were both set with Eq. (\ref{UL_formula}) and via simulation studies conducted with EOBNRv2 waveforms.

\begin{figure*}[htp]
 \centering 
 \subfigure[]{\includegraphics{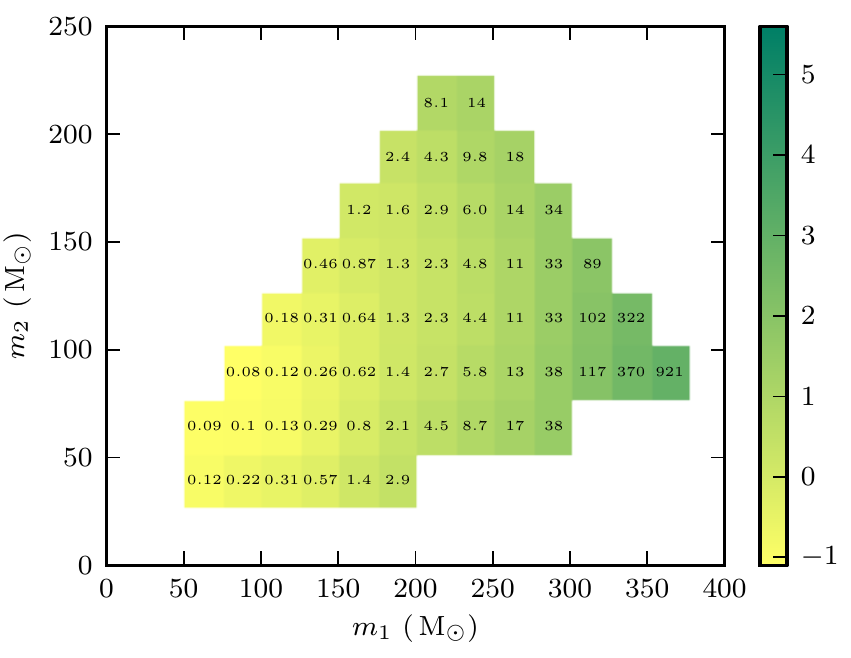}}
 \hspace{3mm}
 \subfigure[]{\includegraphics{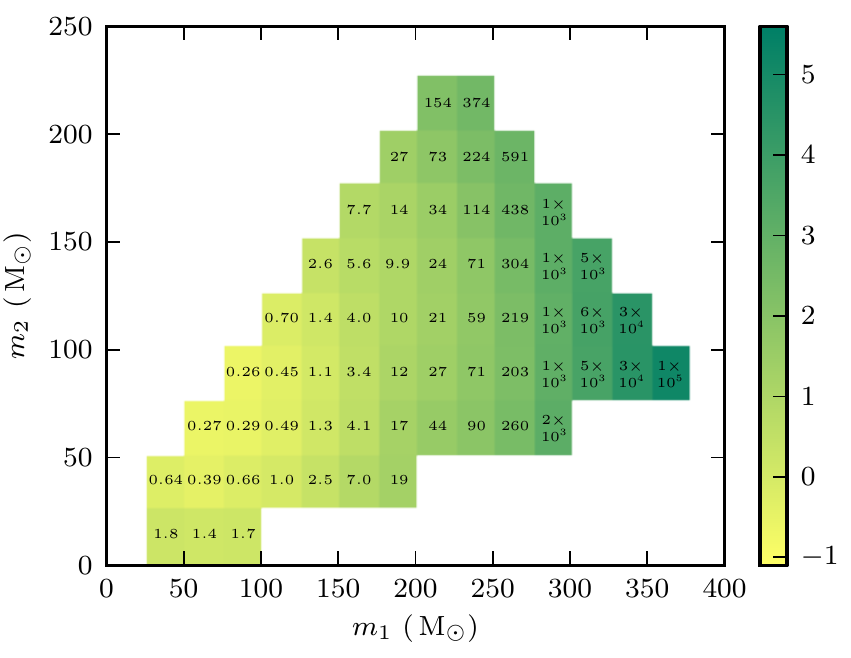}}
  \caption{(color online). (a) S5-VSR1 upper limits in $\text{Mpc}^{-3} \, \text{Myr}^{-1}$ on the merger-rate density of nonspinning IMBHBs as a function of the companion masses. (b) S6-VSR2/3 upper limits in $\text{Mpc}^{-3} \, \text{Myr}^{-1}$ on the merger-rate density of nonspinning IMBHBs as a function of the companion masses. The upper limits were calculated with EOBNRv2 waveforms and were not corrected to account for the uncertainties on the search range. The color scale expresses the upper limits as powers of $10 \,$.} 
 \label{S5_S6_UL_no_systematics}
\end{figure*}

\begin{figure*}[htp]
 \centering
 \subfigure[]{\includegraphics{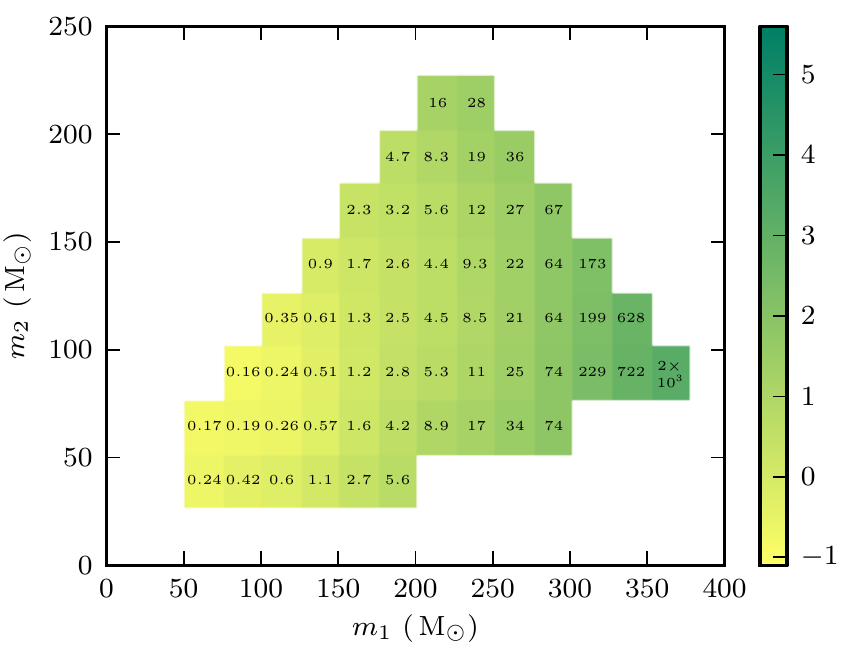}}
 \hspace{3mm}
 \subfigure[]{\includegraphics{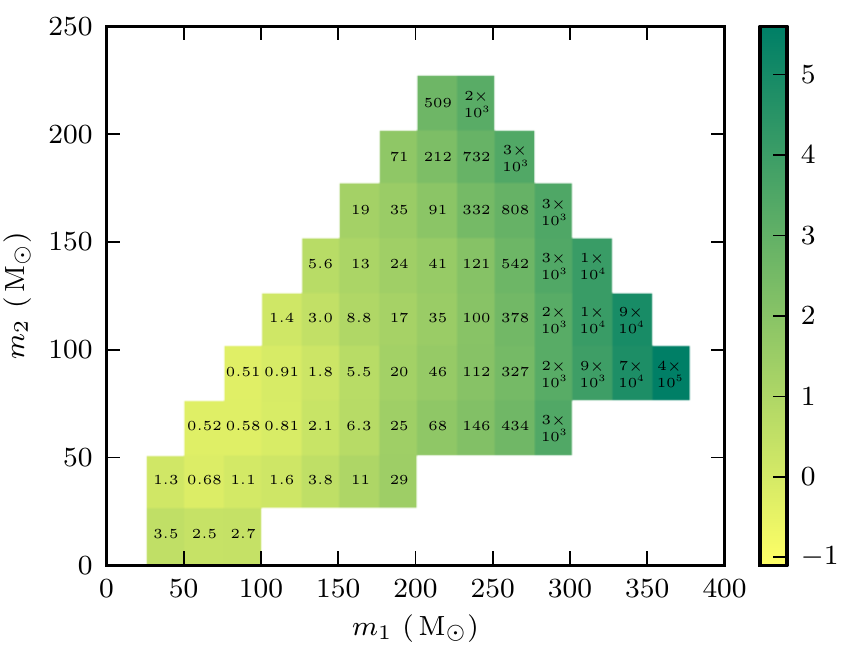}}  
  \caption{(color online). (a) S5-VSR1 upper limits in $\text{Mpc}^{-3} \, \text{Myr}^{-1}$ on the merger-rate density of nonspinning IMBHBs as a function of the companion masses. (b) S6-VSR2/3 upper limits in $\text{Mpc}^{-3} \, \text{Myr}^{-1}$ on the merger-rate density of nonspinning IMBHBs as a function of the companion masses. The upper limits were calculated with EOBNRv2 waveforms and were corrected to account for the uncertainties on the search range. The color scale expresses the upper limits as powers of $10 \,$.} 
 \label{S5_S6_UL_systematics}
\end{figure*}

The S5-VSR1 and S6-VSR2/3 upper limits, calculated without accounting for the uncertainties on the search range discussed in Sec. \ref{impact_of_systematics}$\,$, are shown in Fig. \ref{S5_S6_UL_no_systematics}$\,$. The S5-VSR2 upper limits are more stringent than the S6-VSR2/3 result over the whole investigated parameter space. This was mostly due to the longer observation time analyzed by the S5-VSR1 search compared to the S6-VSR2/3 analysis ($\sim 0.82$ and $\sim 0.33$ yr, respectively [see \cite{Abadie4} and Table \ref{observation_time}$\,$]), the main origin of this difference relying on the longer duration of the S5-VSR1 science run. Aside from the shorter observation time, the lower relevance of the S6-VSR2/3 search was also due to the fact that, during the S6a-VSR2 and S6b-VSR2 epochs, i.e., during more than half of the accumulated H1L1V1 observation time, the threefold configuration showed lower sensitivity compared to S6d-VSR3. This limited the time-volume product $\nu$ collected by the H1L1V1 network. Above $\sim 200 \ \mbox{M}_{\odot}$, the larger discrepancy between the S5-VSR1 and S6-VSR2/3 measures originated also from the hardware components installed at the LIGO facilities after the end of the S5 run \cite{Christensen}. The new components increased the LIGO sensitivity over most of the sensitive band, but also introduced extra noise sources at frequencies below $\sim 60$ Hz.

To calculate conservative, combined upper limits, the S5-VSR1 and S6-VSR2/3 constraints were corrected for the uncertainties on the search range as follows. Hereafter, we will denote the search range with $R_{\text{eff}}$ (for ``effective radius,'' consistently with the notation used in \cite{Abadie4}) and a tilde will denote the observables rescaled to account for the uncertainties. 

First, the $\tilde{R}_{\text{eff}}$ were calculated. The S5-VSR1 $\tilde{R}_{\text{eff}}$ were computed by rescaling the $R_{\text{eff}}$ by the overall uncertainty, equal to $20 \%$ \cite{Abadie4}. The overall S5-VSR1 uncertainty was calculated by summing in quadrature the calibration, waveform and statistical uncertainties. The S6-VSR2/3 $R_{\text{eff}}$ were adjusted to account for the waveform systematics as outlined in Sec. \ref{impact_of_systematics}$ \,$, and subsequently rescaled by the sum in quadrature of the calibration and statistical uncertainties. 

Second, the S5-VSR1 and S6-VSR2/3 $\tilde{V}_{\text{vis}}$ and $\tilde{\nu}$ were calculated, starting from the $\tilde{R}_{\text{eff}} \,$. The associated upper limits, corrected for the uncertainties on $R_{\text{eff}} \,$, were finally computed as
\begin{equation}
 \tilde{\mathcal{R}}_{90 \%} = \frac{2.3}{\tilde{\nu}(\text{FAD}^*)} \ . 
\end{equation}
Here $\text{FAD}^*$ is the FAD threshold at which the S5-VSR1 and S6-VSR2/3 searches have been compared and combined (see Sec. \ref{loudest_event}$\,$). 

The separate S5-VSR1 and S6-VSR2/3 $\tilde{\mathcal{R}}_{90 \%}$ are shown in Fig. \ref{S5_S6_UL_systematics}$ \,$. Comparing the S5-VSR1 plot in Fig. \ref{S5_S6_UL_systematics} to the combined upper limits reported in Fig. \ref{S5_S6_UL} shows that the S6-VSR2/3 contribution decreased the best S5-VSR1 upper limit by $\sim 25 \%$ (from $\sim 0.16$ to $\sim 0.12 \ \mbox{Mpc}^{-3} \, \mbox{Myr}^{-1}$). 

The procedure adopted in this paper to conservatively account for the uncertainties on the search sensitivity differs from the approach followed for the S5-VSR1 analysis. For the S5-VSR1 analysis, the $20 \%$ uncertainty on $R_{\text{eff}}$ was translated into an overall $60 \%$ uncertainty on $V_{\text{vis}}$, which was included in the upper limit calculation by rescaling Eq. (\ref{UL_formula}) by the same amount:
\begin{equation}
 \tilde{\mathcal{R}}_{90 \%} = 1.6 \times \frac{2.3}{\nu(\text{FAD}^*)} \ .
\end{equation}
This led to a less conservative upper limit (compare the S5-VSR1 upper limits in \cite{Abadie4} to the result in Fig. \ref{S5_S6_UL_systematics}). In the most sensitive mass bin, the S5-VSR1 upper limit calculated with the procedure adopted in this Appendix is larger (less stringent) by $\sim 20 \%$ compared to the previous result ($0.16 \ \text{Mpc}^{-3} \ \text{Myr}^{-1}$ rather than $0.13 \ \text{Mpc}^{-3} \ \text{Myr}^{-1}$). The strategy followed for the S5-VSR1 analysis and the formalism outlined in this Appendix provide comparable results only in the case of small fractional uncertainties $\delta$ on the $R_{\text{eff}}$, when $1/(\nu (1-\delta)^3) \approx (1+3\delta)/\nu$. Note that the $\sim 20 \%$ difference between the two S5-VSR1 procedures is comparable to the $\sim 25 \%$ contribution to the combined upper limits from the S6-VSR2/3 search. Thus, in the most sensitive region of the parameter space, the combined upper limits presented in this paper are comparable to the S5-VSR1 upper limits reported in \cite{Abadie4}.

%% file: IMBH_paper.bbl
%Merlin.mbs v4.21 2009-07-09.
\begin{thebibliography}{10}%
\makeatletter
\providecommand \@ifxundefined [1]{%
 \ifx #1\undefined \expandafter \@firstoftwo
 \else \expandafter \@secondoftwo
\fi
}%
\providecommand \@ifnum [1]{%
 \ifnum #1\expandafter \@firstoftwo
 \else \expandafter \@secondoftwo
\fi
}%
\providecommand \enquote [1]{``#1''}%
\providecommand \bibnamefont  [1]{#1}%
\providecommand \bibfnamefont [1]{#1}%
\providecommand \citenamefont [1]{#1}%
\providecommand\href[0]{\@sanitize\@href}%
\providecommand\@href[1]{\endgroup\@@startlink{#1}\endgroup\@@href}%
\providecommand\@@href[1]{#1\@@endlink}%
\providecommand \@sanitize [0]{\begingroup\catcode`\&12\catcode`\#12\relax}%
\@ifxundefined \pdfoutput {\@firstoftwo}{%
 \@ifnum{\z@=\pdfoutput}{\@firstoftwo}{\@secondoftwo}%
}{%
 \providecommand\@@startlink[1]{\leavevmode\special{html:<a href="#1">}}%
 \providecommand\@@endlink[0]{\special{html:</a>}}%
}{%
 \providecommand\@@startlink[1]{%
  \leavevmode
  \pdfstartlink
   attr{/Border[0 0 1 ]/H/I/C[0 1 1]}%
   user{/Subtype/Link/A<</Type/Action/S/URI/URI(#1)>>}%
  \relax
 }%
 \providecommand\@@endlink[0]{\pdfendlink}%
}%
\providecommand \url  [0]{\begingroup\@sanitize \@url }%
\providecommand \@url [1]{\endgroup\@href {#1}{\urlprefix}}%
\providecommand \urlprefix [0]{URL }%
\providecommand \Eprint[0]{\href }%
\@ifxundefined \urlstyle {%
  \providecommand \doi [1]{doi:\discretionary{}{}{}#1}%
}{%
  \providecommand \doi [0]{doi:\discretionary{}{}{}\begingroup
  \urlstyle{rm}\Url }%
}%
\providecommand \doibase [0]{http://dx.doi.org/}%
\providecommand \Doi[1]{\href{\doibase#1}}%
\providecommand \bibAnnote [3]{%
  \BibitemShut{#1}%
  \begin{quotation}\noindent
    \textsc{Key:}\ #2\\\textsc{Annotation:}\ #3%
  \end{quotation}%
}%
\providecommand \bibAnnoteFile [2]{%
  \IfFileExists{#2}{\bibAnnote {#1} {#2} {\input{#2}}}{}%
}%
\providecommand \typeout [0]{\immediate \write \m@ne }%
\providecommand \selectlanguage [0]{\@gobble}%
\providecommand \bibinfo [0]{\@secondoftwo}%
\providecommand \bibfield [0]{\@secondoftwo}%
\providecommand \translation [1]{[#1]}%
\providecommand \BibitemOpen[0]{}%
\providecommand \bibitemStop [0]{}%
\providecommand \bibitemNoStop [0]{.\EOS\space}%
\providecommand \EOS [0]{\spacefactor3000\relax}%
\providecommand \BibitemShut [1]{\csname bibitem#1\endcsname}%
%</preamble>
\bibitem{Coleman_Miller1}%
  \BibitemOpen
  \bibfield{author}{%
  \bibinfo {author} {\bibfnamefont{M.~Coleman }\ \bibnamefont{Miller}}\ and\
  \bibinfo {author} {\bibfnamefont{E.~J.~M. }\ \bibnamefont{Colbert}},\ }%
  \bibfield{journal}{%
  \bibinfo {journal} {Int. J. Mod. Phys.}\ }%
  \textbf{\bibinfo {volume} {D13}},\ \bibinfo {pages} {1} (\bibinfo {year}
  {2004})%
  \bibAnnoteFile{NoStop}{Coleman_Miller1}%
\bibitem{van_der_Marel}%
  \BibitemOpen
  \bibfield{author}{%
  \bibinfo {author} {\bibfnamefont{R.~P. }\ \bibnamefont{van~der Marel}},\ }%
  in\ \emph{\bibinfo {booktitle} {Coevolution of Black holes and Galaxies}},\
  \bibinfo {editor} {edited by\ \bibinfo {editor} {\bibfnamefont{L.~}\
  \bibnamefont{Ho}}}\ (\bibinfo {publisher} {Cambridge University Press,
  Cambridge, England},\ \bibinfo {year} {2004})%
  \bibAnnoteFile{NoStop}{van_der_Marel}%
\bibitem{Sesana}%
  \BibitemOpen
  \bibfield{author}{%
  \bibinfo {author} {\bibfnamefont{A.~}\ \bibnamefont{Sesana}},\ }%
  \bibfield{journal}{%
  \bibinfo {journal} {Adv. Astron.}\ }%
  \textbf{\bibinfo {volume} {2012}},\ \bibinfo {pages} {805402} (\bibinfo
  {year} {2012})%
  \bibAnnoteFile{NoStop}{Sesana}%
\bibitem{Volonteri1}%
  \BibitemOpen
  \bibfield{author}{%
  \bibinfo {author} {\bibfnamefont{M.~}\ \bibnamefont{Volonteri}}\ and\
  \bibinfo {author} {\bibfnamefont{J.~}\ \bibnamefont{Bellovary}},\ }%
  \bibfield{journal}{%
  \bibinfo {journal} {Rep. Prog. Phys.}\ }%
  \textbf{\bibinfo {volume} {75}},\ \bibinfo {pages} {124901} (\bibinfo {year}
  {2012})%
  \bibAnnoteFile{NoStop}{Volonteri1}%
\bibitem{Bash}%
  \BibitemOpen
  \bibfield{author}{%
  \bibinfo {author} {\bibfnamefont{F.~N.~Bash }\ \bibnamefont{\textit{et
  al.}}},\ }%
  \bibfield{journal}{%
  \bibinfo {journal} {Astronom. J.}\ }%
  \textbf{\bibinfo {volume} {135}},\ \bibinfo {pages} {182} (\bibinfo {year}
  {2008})%
  \bibAnnoteFile{NoStop}{Bash}%
\bibitem{Ferraro}%
  \BibitemOpen
  \bibfield{author}{%
  \bibinfo {author} {\bibfnamefont{F.~R.~Ferraro }\ \bibnamefont{\textit{et
  al.}}},\ }%
  \bibfield{journal}{%
  \bibinfo {journal} {Astrophys. J.}\ }%
  \textbf{\bibinfo {volume} {595}},\ \bibinfo {pages} {179} (\bibinfo {year}
  {2003})%
  \bibAnnoteFile{NoStop}{Ferraro}%
\bibitem{Gebhardt}%
  \BibitemOpen
  \bibfield{author}{%
  \bibinfo {author} {\bibfnamefont{K.~Gebhardt }\ \bibnamefont{\textit{et
  al.}}},\ }%
  \bibfield{journal}{%
  \bibinfo {journal} {Astrophys. J.}\ }%
  \textbf{\bibinfo {volume} {634}},\ \bibinfo {pages} {1093} (\bibinfo {year}
  {2005})%
  \bibAnnoteFile{NoStop}{Gebhardt}%
\bibitem{Noyola}%
  \BibitemOpen
  \bibfield{author}{%
  \bibinfo {author} {\bibfnamefont{E.~Noyola }\ \bibnamefont{\textit{et
  al.}}},\ }%
  \bibfield{journal}{%
  \bibinfo {journal} {Astrophys. J.}\ }%
  \textbf{\bibinfo {volume} {676}},\ \bibinfo {pages} {1008} (\bibinfo {year}
  {2008})%
  \bibAnnoteFile{NoStop}{Noyola}%
\bibitem{Trenti}%
  \BibitemOpen
  \bibfield{author}{%
  \bibinfo {author} {\bibfnamefont{M.~}\ \bibnamefont{Trenti}}\ }%
  \Eprint{http://arxiv.org/abs/astro-ph/0612040}{arXiv:astro-ph/0612040}%
  \bibAnnoteFile{NoStop}{Trenti}%
\bibitem{van_den_Bosch}%
  \BibitemOpen
  \bibfield{author}{%
  \bibinfo {author} {\bibfnamefont{R.~}\ \bibnamefont{van~den Bosch~\textit{et
  al.}}},\ }%
  \bibfield{journal}{%
  \bibinfo {journal} {Astrophys. J.}\ }%
  \textbf{\bibinfo {volume} {641}},\ \bibinfo {pages} {852} (\bibinfo {year}
  {2006})%
  \bibAnnoteFile{NoStop}{van_den_Bosch}%
\bibitem{Farrell}%
  \BibitemOpen
  \bibfield{author}{%
  \bibinfo {author} {\bibfnamefont{S.~A.~Farrell }\ \bibnamefont{\textit{et
  al.}}},\ }%
  \bibfield{journal}{%
  \bibinfo {journal} {Nature}\ }%
  \textbf{\bibinfo {volume} {460}},\ \bibinfo {pages} {73} (\bibinfo {year}
  {2009})%
  \bibAnnoteFile{NoStop}{Farrell}%
\bibitem{Farrell2}%
  \BibitemOpen
  \bibfield{author}{%
  \bibinfo {author} {\bibfnamefont{S.~A.~Farrell }\ \bibnamefont{\textit{et
  al.}}},\ }%
  \bibfield{journal}{%
  \bibinfo {journal} {Mon. Not. R. Astronom. Soc.}\ }%
  \textbf{\bibinfo {volume} {437}},\ \bibinfo {pages} {1208} (\bibinfo {year}
  {2013})%
  \bibAnnoteFile{NoStop}{Farrell2}%
\bibitem{Kaaret}%
  \BibitemOpen
  \bibfield{author}{%
  \bibinfo {author} {\bibfnamefont{P.~Kaaret }\ \bibnamefont{\textit{et
  al.}}},\ }%
  \bibfield{journal}{%
  \bibinfo {journal} {Mon. Not. R. Astron. Soc.}\ }%
  \textbf{\bibinfo {volume} {351}},\ \bibinfo {pages} {L83} (\bibinfo {year}
  {2004})%
  \bibAnnoteFile{NoStop}{Kaaret}%
\bibitem{Kajava}%
  \BibitemOpen
  \bibfield{author}{%
  \bibinfo {author} {\bibfnamefont{J.~J.~E. }\ \bibnamefont{Kajava}}\ and\
  \bibinfo {author} {\bibfnamefont{J.~}\ \bibnamefont{Poutanen}},\ }%
  \bibfield{journal}{%
  \bibinfo {journal} {Mon. Not. R. Astron. Soc.}\ }%
  \textbf{\bibinfo {volume} {398}},\ \bibinfo {pages} {1450} (\bibinfo {year}
  {2009})%
  \bibAnnoteFile{NoStop}{Kajava}%
\bibitem{Strohmayer1}%
  \BibitemOpen
  \bibfield{author}{%
  \bibinfo {author} {\bibfnamefont{T.~E. }\ \bibnamefont{Strohmayer}}\ and\
  \bibinfo {author} {\bibfnamefont{R.~F. }\ \bibnamefont{Mushotzky}},\ }%
  \bibfield{journal}{%
  \bibinfo {journal} {Astrophys. J.}\ }%
  \textbf{\bibinfo {volume} {586}},\ \bibinfo {pages} {L61} (\bibinfo {year}
  {2003})%
  \bibAnnoteFile{NoStop}{Strohmayer1}%
\bibitem{Strohmayer2}%
  \BibitemOpen
  \bibfield{author}{%
  \bibinfo {author} {\bibfnamefont{T.~E.~Strohmayer }\ \bibnamefont{\textit{et
  al.}}},\ }%
  \bibfield{journal}{%
  \bibinfo {journal} {Astrophys. J.}\ }%
  \textbf{\bibinfo {volume} {660}},\ \bibinfo {pages} {580} (\bibinfo {year}
  {2007})%
  \bibAnnoteFile{NoStop}{Strohmayer2}%
\bibitem{Vierdayanti}%
  \BibitemOpen
  \bibfield{author}{%
  \bibinfo {author} {\bibfnamefont{K.~Vierdayanti }\ \bibnamefont{\textit{et
  al.}}},\ }%
  \bibfield{journal}{%
  \bibinfo {journal} {Publ. Astron. Soc. Jpn.}\ }%
  \textbf{\bibinfo {volume} {60}},\ \bibinfo {pages} {653} (\bibinfo {year}
  {2008})%
  \bibAnnoteFile{NoStop}{Vierdayanti}%
\bibitem{Abbott1}%
  \BibitemOpen
  \bibfield{author}{%
  \bibinfo {author} {\bibfnamefont{B.~P.~Abbott }\ \bibnamefont{\textit{et
  al.}}} (\bibinfo {collaboration} {LIGO Scientific Collaboration}),\ }%
  \bibfield{journal}{%
  \bibinfo {journal} {Rep. Prog. Phys.}\ }%
  \textbf{\bibinfo {volume} {72}},\ \bibinfo {pages} {076901} (\bibinfo {year}
  {2009})%
  \bibAnnoteFile{NoStop}{Abbott1}%
\bibitem{Accadia}%
  \BibitemOpen
  \bibfield{author}{%
  \bibinfo {author} {\bibfnamefont{T.~Accadia }\ \bibnamefont{\textit{et al.}}}
  (\bibinfo {collaboration} {Virgo Collaboration}),\ }%
  \bibfield{journal}{%
  \bibinfo {journal} {JINST}\ }%
  \textbf{\bibinfo {volume} {7}},\ \bibinfo {pages} {P03012} (\bibinfo {year}
  {2012})%
  \bibAnnoteFile{NoStop}{Accadia}%
\bibitem{Abbott2}%
  \BibitemOpen
  \bibfield{author}{%
  \bibinfo {author} {\bibfnamefont{B.~P.~Abbott }\ \bibnamefont{\textit{et
  al.}}} (\bibinfo {collaboration} {LIGO Scientific Collaboration}),\ }%
  \bibfield{journal}{%
  \bibinfo {journal} {Phys. Rev. D}\ }%
  \textbf{\bibinfo {volume} {79}},\ \bibinfo {pages} {122001} (\bibinfo {year}
  {2009})%
  \bibAnnoteFile{NoStop}{Abbott2}%
\bibitem{Abbott3}%
  \BibitemOpen
  \bibfield{author}{%
  \bibinfo {author} {\bibfnamefont{B.~P.~Abbott }\ \bibnamefont{\textit{et
  al.}}} (\bibinfo {collaboration} {LIGO Scientific Collaboration}),\ }%
  \bibfield{journal}{%
  \bibinfo {journal} {Phys. Rev. D}\ }%
  \textbf{\bibinfo {volume} {80}},\ \bibinfo {pages} {047101} (\bibinfo {year}
  {2009})%
  \bibAnnoteFile{NoStop}{Abbott3}%
\bibitem{Abadie1}%
  \BibitemOpen
  \bibfield{author}{%
  \bibinfo {author} {\bibfnamefont{J.~Abadie }\ \bibnamefont{\textit{et al.}}}
  (\bibinfo {collaboration} {LIGO Scientific Collaboration and Virgo
  Collaboration}),\ }%
  \bibfield{journal}{%
  \bibinfo {journal} {Phys. Rev. D}\ }%
  \textbf{\bibinfo {volume} {82}},\ \bibinfo {pages} {102001} (\bibinfo {year}
  {2010})%
  \bibAnnoteFile{NoStop}{Abadie1}%
\bibitem{Abadie2}%
  \BibitemOpen
  \bibfield{author}{%
  \bibinfo {author} {\bibfnamefont{J.~Abadie }\ \bibnamefont{\textit{et al.}}}
  (\bibinfo {collaboration} {LIGO Scientific Collaboration and Virgo
  Collaboration}),\ }%
  \bibfield{journal}{%
  \bibinfo {journal} {Phys. Rev. D}\ }%
  \textbf{\bibinfo {volume} {85}},\ \bibinfo {pages} {082002} (\bibinfo {year}
  {2012})%
  \bibAnnoteFile{NoStop}{Abadie2}%
\bibitem{Abadie3}%
  \BibitemOpen
  \bibfield{author}{%
  \bibinfo {author} {\bibfnamefont{J.~Abadie }\ \bibnamefont{\textit{et al.}}}
  (\bibinfo {collaboration} {LIGO Scientific Collaboration and Virgo
  Collaboration}),\ }%
  \bibfield{journal}{%
  \bibinfo {journal} {Phys. Rev. D}\ }%
  \textbf{\bibinfo {volume} {83}},\ \bibinfo {pages} {122005} (\bibinfo {year}
  {2011})%
  \bibAnnoteFile{NoStop}{Abadie3}%
\bibitem{Aasi1}%
  \BibitemOpen
  \bibfield{author}{%
  \bibinfo {author} {\bibfnamefont{J.~Aasi }\ \bibnamefont{\textit{et al.}}}
  (\bibinfo {collaboration} {LIGO Scientific Collaboration and Virgo
  Collaboration}),\ }%
  \bibfield{journal}{%
  \bibinfo {journal} {Phys. Rev. D}\ }%
  \textbf{\bibinfo {volume} {87}},\ \bibinfo {pages} {022002} (\bibinfo {year}
  {2013})%
  \bibAnnoteFile{NoStop}{Aasi1}%
\bibitem{Abbott4}%
  \BibitemOpen
  \bibfield{author}{%
  \bibinfo {author} {\bibfnamefont{B.~P.~Abbott }\ \bibnamefont{\textit{et
  al.}}} (\bibinfo {collaboration} {LIGO Scientific Collaboration}),\ }%
  \bibfield{journal}{%
  \bibinfo {journal} {Phys. Rev. D}\ }%
  \textbf{\bibinfo {volume} {80}},\ \bibinfo {pages} {062001} (\bibinfo {year}
  {2009})%
  \bibAnnoteFile{NoStop}{Abbott4}%
\bibitem{Aasi4}%
  \BibitemOpen
  \bibfield{author}{%
  \bibinfo {author} {\bibfnamefont{J.~Aasi }\ \bibnamefont{\textit{et al.}}}
  (\bibinfo {collaboration} {LIGO Scientific Collaboration and Virgo
  Collaboration}),\ }%
  \bibfield{journal}{%
  \bibinfo {journal} {Phys. Rev. D}\ }%
  \textbf{\bibinfo {volume} {89}},\ \bibinfo {pages} {102006} (\bibinfo {year}
  {2014})%
  \bibAnnoteFile{NoStop}{Aasi4}%
\bibitem{Anderson}%
  \BibitemOpen
  \bibfield{author}{%
  \bibinfo {author} {\bibfnamefont{W.~G.~Anderson }\ \bibnamefont{\textit{et
  al.}}},\ }%
  \bibfield{journal}{%
  \bibinfo {journal} {Phys. Rev. D}\ }%
  \textbf{\bibinfo {volume} {63}},\ \bibinfo {pages} {042003} (\bibinfo {year}
  {2001})%
  \bibAnnoteFile{NoStop}{Anderson}%
\bibitem{Mohapatra}%
  \BibitemOpen
  \bibfield{author}{%
  \bibinfo {author} {\bibfnamefont{S.~Mohapatra }\ \bibnamefont{\textit{et
  al.}}}\ }%
  \Eprint{http://arxiv.org/abs/1405.6589}{arXiv:1405.6589 [gr-qc]}%
  \bibAnnoteFile{NoStop}{Mohapatra}%
\bibitem{Hannam2}%
  \BibitemOpen
  \bibfield{author}{%
  \bibinfo {author} {\bibfnamefont{M.~}\ \bibnamefont{Hannam}}\ }%
  \Eprint{http://arxiv.org/abs/1312.3641}{arXiv:1312.3641 [gr-qc]}%
  \bibAnnoteFile{NoStop}{Hannam2}%
\bibitem{Hannam:2013waveform}%
  \BibitemOpen
  \bibfield{author}{%
  \bibinfo {author} {\bibfnamefont{M.~Hannam }\ \bibnamefont{\textit{et al.}}}\
  }%
  \Eprint{http://arxiv.org/abs/1308.3271}{arXiv:1308.3271 [gr-qc]}%
  \bibAnnoteFile{NoStop}{Hannam:2013waveform}%
\bibitem{Pan2}%
  \BibitemOpen
  \bibfield{author}{%
  \bibinfo {author} {\bibfnamefont{Y.~Pan }\ \bibnamefont{\textit{et al.}}},\
  }%
  \bibfield{journal}{%
  \bibinfo {journal} {Phys. Rev. D}\ }%
  \textbf{\bibinfo {volume} {89}},\ \bibinfo {pages} {084006} (\bibinfo {year}
  {2014})%
  \bibAnnoteFile{NoStop}{Pan2}%
\bibitem{Taracchini:2013}%
  \BibitemOpen
  \bibfield{author}{%
  \bibinfo {author} {\bibfnamefont{A.~Taracchini }\ \bibnamefont{\textit{et
  al.}}},\ }%
  \bibfield{journal}{%
  \bibinfo {journal} {Phys. Rev. D}\ }%
  \textbf{\bibinfo {volume} {89}},\ \bibinfo {pages} {061502} (\bibinfo {year}
  {2014})%
  \bibAnnoteFile{NoStop}{Taracchini:2013}%
\bibitem{Sturani}%
  \BibitemOpen
  \bibfield{author}{%
  \bibinfo {author} {\bibfnamefont{R.~Sturani }\ \bibnamefont{\textit{et
  al.}}},\ }%
  \bibfield{journal}{%
  \bibinfo {journal} {J. Phys.: Conf. Ser.}\ }%
  \textbf{\bibinfo {volume} {243}},\ \bibinfo {pages} {012007} (\bibinfo {year}
  {2010})%
  \bibAnnoteFile{NoStop}{Sturani}%
\bibitem{Klimenko1}%
  \BibitemOpen
  \bibfield{author}{%
  \bibinfo {author} {\bibfnamefont{S.~Klimenko }\ \bibnamefont{\textit{et
  al.}}},\ }%
  \bibfield{journal}{%
  \bibinfo {journal} {Class. Quantum Grav.}\ }%
  \textbf{\bibinfo {volume} {25}},\ \bibinfo {pages} {114029} (\bibinfo {year}
  {2008})%
  \bibAnnoteFile{NoStop}{Klimenko1}%
\bibitem{Abadie4}%
  \BibitemOpen
  \bibfield{author}{%
  \bibinfo {author} {\bibfnamefont{J.~Abadie }\ \bibnamefont{\textit{et al.}}}
  (\bibinfo {collaboration} {LIGO Scientific Collaboration and Virgo
  Collaboration}),\ }%
  \bibfield{journal}{%
  \bibinfo {journal} {Phys. Rev. D}\ }%
  \textbf{\bibinfo {volume} {85}},\ \bibinfo {pages} {102004} (\bibinfo {year}
  {2012})%
  \bibAnnoteFile{NoStop}{Abadie4}%
\bibitem{Slutsky}%
  \BibitemOpen
  \bibfield{author}{%
  \bibinfo {author} {\bibfnamefont{J.~Slutsky }\ \bibnamefont{\textit{et
  al.}}},\ }%
  \bibfield{journal}{%
  \bibinfo {journal} {Class. Quantum Grav.}\ }%
  \textbf{\bibinfo {volume} {27}},\ \bibinfo {pages} {165023} (\bibinfo {year}
  {2010})%
  \bibAnnoteFile{NoStop}{Slutsky}%
\bibitem{Aasi3}%
  \BibitemOpen
  \bibfield{author}{%
  \bibinfo {author} {\bibfnamefont{J.~Aasi }\ \bibnamefont{\textit{et al.}}}
  (\bibinfo {collaboration} {LIGO Scientific Collaboration and Virgo
  Collaboration}),\ }%
  \bibfield{journal}{%
  \bibinfo {journal} {Class. Quantum Grav.}\ }%
  \textbf{\bibinfo {volume} {29}},\ \bibinfo {pages} {155002} (\bibinfo {year}
  {2012})%
  \bibAnnoteFile{NoStop}{Aasi3}%
\bibitem{McIver}%
  \BibitemOpen
  \bibfield{author}{%
  \bibinfo {author} {\bibfnamefont{J.~}\ \bibnamefont{McIver}},\ }%
  \bibfield{journal}{%
  \bibinfo {journal} {Class. Quantum Grav.}\ }%
  \textbf{\bibinfo {volume} {29}},\ \bibinfo {pages} {124010} (\bibinfo {year}
  {2012})%
  \bibAnnoteFile{NoStop}{McIver}%
\bibitem{Smith}%
  \BibitemOpen
  \bibfield{author}{%
  \bibinfo {author} {\bibfnamefont{J.~R.~Smith }\ \bibnamefont{\textit{et
  al.}}},\ }%
  \bibfield{journal}{%
  \bibinfo {journal} {Class. Quantum Grav.}\ }%
  \textbf{\bibinfo {volume} {28}},\ \bibinfo {pages} {235005} (\bibinfo {year}
  {2011})%
  \bibAnnoteFile{NoStop}{Smith}%
\bibitem{Klimenko2}%
  \BibitemOpen
  \bibfield{author}{%
  \bibinfo {author} {\bibfnamefont{S.~}\ \bibnamefont{Klimenko}}\ and\ \bibinfo
  {author} {\bibfnamefont{G.~}\ \bibnamefont{Mitselmakher}},\ }%
  \bibfield{journal}{%
  \bibinfo {journal} {Class. Quantum Grav.}\ }%
  \textbf{\bibinfo {volume} {21}},\ \bibinfo {pages} {S1819} (\bibinfo {year}
  {2004})%
  \bibAnnoteFile{NoStop}{Klimenko2}%
\bibitem{Klimenko3}%
  \BibitemOpen
  \bibfield{author}{%
  \bibinfo {author} {\bibfnamefont{S.~Klimenko }\ \bibnamefont{\textit{et
  al.}}},\ }%
  \bibfield{journal}{%
  \bibinfo {journal} {Phys. Rev. D}\ }%
  \textbf{\bibinfo {volume} {72}},\ \bibinfo {pages} {122002} (\bibinfo {year}
  {2005})%
  \bibAnnoteFile{NoStop}{Klimenko3}%
\bibitem{Klimenko4}%
  \BibitemOpen
  \bibfield{author}{%
  \bibinfo {author} {\bibfnamefont{S.~Klimenko }\ \bibnamefont{\textit{et
  al.}}},\ }%
  \bibfield{journal}{%
  \bibinfo {journal} {J. Phys.: Conf. Ser.}\ }%
  \textbf{\bibinfo {volume} {32}},\ \bibinfo {pages} {12} (\bibinfo {year}
  {2006})%
  \bibAnnoteFile{NoStop}{Klimenko4}%
\bibitem{Pankow1}%
  \BibitemOpen
  \bibfield{author}{%
  \bibinfo {author} {\bibfnamefont{C.~Pankow }\ \bibnamefont{\textit{et
  al.}}},\ }%
  \bibfield{journal}{%
  \bibinfo {journal} {Class. Quantum Grav.}\ }%
  \textbf{\bibinfo {volume} {26}},\ \bibinfo {pages} {204004} (\bibinfo {year}
  {2009})%
  \bibAnnoteFile{NoStop}{Pankow1}%
\bibitem{Pan}%
  \BibitemOpen
  \bibfield{author}{%
  \bibinfo {author} {\bibfnamefont{Y.~Pan }\ \bibnamefont{\textit{et al.}}},\
  }%
  \bibfield{journal}{%
  \bibinfo {journal} {Phys. Rev. D}\ }%
  \textbf{\bibinfo {volume} {84}},\ \bibinfo {pages} {124052} (\bibinfo {year}
  {2011})%
  \bibAnnoteFile{NoStop}{Pan}%
\bibitem{Ajith2}%
  \BibitemOpen
  \bibfield{author}{%
  \bibinfo {author} {\bibfnamefont{P.~Ajith }\ \bibnamefont{\textit{et al.}}},\
  }%
  \bibfield{journal}{%
  \bibinfo {journal} {Phys. Rev. Lett.}\ }%
  \textbf{\bibinfo {volume} {106}},\ \bibinfo {pages} {241101} (\bibinfo {year}
  {2011})%
  \bibAnnoteFile{NoStop}{Ajith2}%
\bibitem{Pankow2}%
  \BibitemOpen
  \bibfield{author}{%
  \bibinfo {author} {\bibfnamefont{C.~}\ \bibnamefont{Pankow}},\ }%
  \bibfield{journal}{%
  \bibinfo {journal} {Ph. D. thesis, University of Florida}}%
   (\bibinfo {year} {2011}),\
  \url{http://uf.catalog.fcla.edu/uf.jsp?st=UF005295299&ix=pm&I=0&V=D&pm=1}%
  \bibAnnoteFile{NoStop}{Pankow2}%
\bibitem{Bartos}%
  \BibitemOpen
  \bibfield{author}{%
  \bibinfo {author} {\bibfnamefont{I.~Bartos }\ \bibnamefont{\textit{et
  al.}}},\ }%
  \bibfield{journal}{%
  \bibinfo {journal} {LIGO Document T1100071-v9}}%
   (\bibinfo {year} {2012}),\
  \url{https://dcc.ligo.org/LIGO-T1100071-v9/public}%
  \bibAnnoteFile{NoStop}{Bartos}%
\bibitem{Accadia4}%
  \BibitemOpen
  \bibfield{author}{%
  \bibinfo {author} {\bibfnamefont{T.~Accadia }\ \bibnamefont{\textit{et al.}}}
  (\bibinfo {collaboration} {Virgo Collaboration})\ }%
  \Eprint{http://arxiv.org/abs/1401.6066}{arXiv:1401.6066 [gr-qc]}%
  \bibAnnoteFile{NoStop}{Accadia4}%
\bibitem{Accadia2}%
  \BibitemOpen
  \bibfield{author}{%
  \bibinfo {author} {\bibfnamefont{T.~Accadia }\ \bibnamefont{\textit{et al.}}}
  (\bibinfo {collaboration} {Virgo Collaboration}),\ }%
  \bibfield{journal}{%
  \bibinfo {journal} {Class. Quantum Grav.}\ }%
  \textbf{\bibinfo {volume} {28}},\ \bibinfo {pages} {025005} (\bibinfo {year}
  {2011})%
  \bibAnnoteFile{NoStop}{Accadia2}%
\bibitem{S6PECBC}%
  \BibitemOpen
  \bibfield{author}{%
  \bibinfo {author} {\bibfnamefont{J.~Aasi }\ \bibnamefont{\textit{et al.}}}
  (\bibinfo {collaboration} {LIGO Scientific Collaboration and Virgo
  Collaboration}),\ }%
  \bibfield{journal}{%
  \bibinfo {journal} {Phys. Rev. D}\ }%
  \textbf{\bibinfo {volume} {88}},\ \bibinfo {pages} {062001} (\bibinfo {year}
  {2013})%
  \bibAnnoteFile{NoStop}{S6PECBC}%
\bibitem{McClintock}%
  \BibitemOpen
  \bibfield{author}{%
  \bibinfo {author} {\bibfnamefont{J.~E.~McClintock }\ \bibnamefont{\textit{et
  al.}}},\ }%
  \bibfield{journal}{%
  \bibinfo {journal} {Class. Quantum Grav.}\ }%
  \textbf{\bibinfo {volume} {28}},\ \bibinfo {pages} {114009} (\bibinfo {year}
  {2011})%
  \bibAnnoteFile{NoStop}{McClintock}%
\bibitem{Reynolds}%
  \BibitemOpen
  \bibfield{author}{%
  \bibinfo {author} {\bibfnamefont{C.~S. }\ \bibnamefont{Reynolds}},\ }%
  \bibfield{journal}{%
  \bibinfo {journal} {Class. Quantum Grav.}\ }%
  \textbf{\bibinfo {volume} {30}},\ \bibinfo {pages} {244004} (\bibinfo {year}
  {2013})%
  \bibAnnoteFile{NoStop}{Reynolds}%
\bibitem{Campanelli}%
  \BibitemOpen
  \bibfield{author}{%
  \bibinfo {author} {\bibfnamefont{M.~Campanelli }\ \bibnamefont{\textit{et
  al.}}},\ }%
  \bibfield{journal}{%
  \bibinfo {journal} {Phys. Rev. D}\ }%
  \textbf{\bibinfo {volume} {74}},\ \bibinfo {pages} {041501} (\bibinfo {year}
  {2006})%
  \bibAnnoteFile{NoStop}{Campanelli}%
\bibitem{Gerosa}%
  \BibitemOpen
  \bibfield{author}{%
  \bibinfo {author} {\bibfnamefont{D.~Gerosa }\ \bibnamefont{\textit{et
  al.}}},\ }%
  \bibfield{journal}{%
  \bibinfo {journal} {Phys. Rev. D}\ }%
  \textbf{\bibinfo {volume} {87}},\ \bibinfo {pages} {104028} (\bibinfo {year}
  {2013})%
  \bibAnnoteFile{NoStop}{Gerosa}%
\bibitem{Apostolatos}%
  \BibitemOpen
  \bibfield{author}{%
  \bibinfo {author} {\bibfnamefont{T.~A.~Apostolatos }\ \bibnamefont{\textit{et
  al.}}},\ }%
  \bibfield{journal}{%
  \bibinfo {journal} {Phys. Rev. D}\ }%
  \textbf{\bibinfo {volume} {49}},\ \bibinfo {pages} {6274} (\bibinfo {year}
  {1994})%
  \bibAnnoteFile{NoStop}{Apostolatos}%
\bibitem{Biswas}%
  \BibitemOpen
  \bibfield{author}{%
  \bibinfo {author} {\bibfnamefont{R.~Biswas }\ \bibnamefont{\textit{et
  al.}}},\ }%
  \bibfield{journal}{%
  \bibinfo {journal} {Class. Quantum Grav.}\ }%
  \textbf{\bibinfo {volume} {26}},\ \bibinfo {pages} {175009} (\bibinfo {year}
  {2009})%
  \bibAnnoteFile{NoStop}{Biswas}%
\bibitem{Sutton}%
  \BibitemOpen
  \bibfield{author}{%
  \bibinfo {author} {\bibfnamefont{P.~J. }\ \bibnamefont{Sutton}},\ }%
  \bibfield{journal}{%
  \bibinfo {journal} {Class. Quantum Grav.}\ }%
  \textbf{\bibinfo {volume} {26}},\ \bibinfo {pages} {245007} (\bibinfo {year}
  {2009})%
  \bibAnnoteFile{NoStop}{Sutton}%
\bibitem{Coleman_Miller3}%
  \BibitemOpen
  \bibfield{author}{%
  \bibinfo {author} {\bibfnamefont{M.~Coleman }\ \bibnamefont{Miller}}\ and\
  \bibinfo {author} {\bibfnamefont{D.~P. }\ \bibnamefont{Hamilton}},\ }%
  \bibfield{journal}{%
  \bibinfo {journal} {Mon. Not. R. Astron. Soc.}\ }%
  \textbf{\bibinfo {volume} {330}},\ \bibinfo {pages} {232} (\bibinfo {year}
  {2002})%
  \bibAnnoteFile{NoStop}{Coleman_Miller3}%
\bibitem{Portegies2}%
  \BibitemOpen
  \bibfield{author}{%
  \bibinfo {author} {\bibfnamefont{S.~F.~Portegies }\ \bibnamefont{Zwart}}\
  and\ \bibinfo {author} {\bibfnamefont{S.~L.~W. }\ \bibnamefont{McMillan}},\
  }%
  \bibfield{journal}{%
  \bibinfo {journal} {Astrophys. J.}\ }%
  \textbf{\bibinfo {volume} {528}},\ \bibinfo {pages} {L17} (\bibinfo {year}
  {2000})%
  \bibAnnoteFile{NoStop}{Portegies2}%
\bibitem{Harry}%
  \BibitemOpen
  \bibfield{author}{%
  \bibinfo {author} {\bibfnamefont{G.~M. }\ \bibnamefont{Harry}} (\bibinfo
  {collaboration} {for the LIGO Scientific Collaboration}),\ }%
  \bibfield{journal}{%
  \bibinfo {journal} {Class. Quantum Grav.}\ }%
  \textbf{\bibinfo {volume} {27}},\ \bibinfo {pages} {084006} (\bibinfo {year}
  {2010})%
  \bibAnnoteFile{NoStop}{Harry}%
\bibitem{Accadia3}%
  \BibitemOpen
  \bibfield{author}{%
  \bibinfo {author} {\bibfnamefont{Virgo }\ \bibnamefont{Collaboration}},\ }%
  \bibfield{journal}{%
  \bibinfo {journal} {Virgo Document VIR-0128A-12}}%
   (\bibinfo {year} {2012}),\ \url{https://tds.ego-gw.it/ql/?c=8940}%
  \bibAnnoteFile{NoStop}{Accadia3}%
\bibitem{Somiya}%
  \BibitemOpen
  \bibfield{author}{%
  \bibinfo {author} {\bibfnamefont{K.~}\ \bibnamefont{Somiya}} (\bibinfo
  {collaboration} {for the KAGRA Collaboration}),\ }%
  \bibfield{journal}{%
  \bibinfo {journal} {Class. Quantum Grav.}\ }%
  \textbf{\bibinfo {volume} {29}},\ \bibinfo {pages} {124007} (\bibinfo {year}
  {2012})%
  \bibAnnoteFile{NoStop}{Somiya}%
\bibitem{Aasi2}%
  \BibitemOpen
  \bibfield{author}{%
  \bibinfo {author} {\bibfnamefont{J.~Aasi }\ \bibnamefont{\textit{et al.}}}
  (\bibinfo {collaboration} {LIGO Scientific Collaboration and Virgo
  Collaboration})\ }%
  \Eprint{http://arxiv.org/abs/1304.0670}{arXiv:1304.0670 [gr-qc]}%
  \bibAnnoteFile{NoStop}{Aasi2}%
\bibitem{Mazzolo2}%
  \BibitemOpen
  \bibfield{author}{%
  \bibinfo {author} {\bibfnamefont{G.~Mazzolo }\ \bibnamefont{\textit{et
  al.}}}\ }%
  \Eprint{http://arxiv.org/abs/1404.7757}{arXiv:1404.7757 [gr-qc]}%
  \bibAnnoteFile{NoStop}{Mazzolo2}%
\bibitem{Christensen}%
  \BibitemOpen
  \bibfield{author}{%
  \bibinfo {author} {\bibfnamefont{N.~}\ \bibnamefont{Christensen}} (\bibinfo
  {collaboration} {for the LIGO Scientific Collaboration and the Virgo
  Collaboration}),\ }%
  \bibfield{journal}{%
  \bibinfo {journal} {Class. Quantum Grav.}\ }%
  \textbf{\bibinfo {volume} {27}},\ \bibinfo {pages} {194010} (\bibinfo {year}
  {2010})%
  \bibAnnoteFile{NoStop}{Christensen}%
\end{thebibliography}%
